\documentclass{ar-1col-S2O}
\usepackage{url}
\usepackage[numbers]{natbib}
\usepackage{xspace}
\usepackage{graphicx}
\usepackage{color}
\usepackage{float}
\usepackage{amssymb}
\usepackage{amsmath}
\usepackage{array}
\usepackage{hyperref}
\usepackage{lmodern}
\usepackage{subfigure}
\usepackage{slashed}
\usepackage{relsize}

\input symbols.tex
            
\jname{Annu. Rev. Nucl. Part. Sci.}
\jvol{76}
\fstpage{119}
\endpage{144}
\jyear{2026}
\doi{10.1146/annurev-nucl-100324-024335}

\begin{document}

\markboth{Echenard and Petrov}{Precision Physics with Muons}

\title{Precision Physics with Muons : A Decade of Theoretical and Experimental Advances}

\author{Bertrand Echenard$^1$ and Alexey A. Petrov$^2$
\affil{$^1$Department of Physics, Mathematics and Astronomy, California Institute of Technology, Pasadena, USA, CA 91125; email: echenard@caltech.edu}
\affil{$^2$Department of Physics and Astronomy, University of South Carolina, USA, SC 29208}
}

\begin{abstract}
The muon has been instrumental in establishing the Standard Model of particle physics and continues to play a key role in exploring the nature of New Physics. A global program is underway to enhance the discovery potential of a wide range of muon probes, with significant increases in sensitivity anticipated over the next decade. In this review, we examine recent experimental advancements in the study of muon decays, the determination of the muon magnetic and electric dipole moments, and the search for charged lepton flavor-violating transitions. We explore the implications for scenarios of physics beyond the Standard Model, focusing on models involving light new particles, such as axions and hidden sectors. Opportunities from novel experimental concepts and proposals for new muon facilities are also discussed.
\end{abstract}

\maketitle

\tableofcontents

\section{INTRODUCTION}
Since its discovery in cosmic rays~\cite{Anderson:1936zz}, the muon has played a central role in establishing the Standard Model (SM) of particle physics. Initially mistaken for the mediator of the nuclear force predicted by Yukawa~\cite{Yukawa:1935xg}, its nature was elucidated through experiments by Conversi, Pancini, and Piccioni in the 1940s~\cite{Conversi:1945qhg, Conversi:1947ig}, which demonstrated that the muon interacts only weakly with matter and must be different from the Yukawa particle (now known as the pion). Subsequent investigations by Hincks \& Pontecorvo~\cite{Hincks:1948vr} revealed no evidence of transitions of a muon into an electron and a photon, excluding the possibility that the muon is an excited state of the electron. By the end of the 1950s, the experimental limit on $\mu \rightarrow e \gamma$ decays was well below the predictions of weak interaction models with a single neutrino. This apparent crisis was resolved by postulating the existence of two neutrino flavors (and incidentally lepton flavor conservation), experimentally confirmed by the discovery of the muon neutrino a few years later~\cite{Danby:1962nd}. In parallel, experiments on parity non-conservation in muon decays during the 1950s yielded the first measurements of its magnetic moment~\cite{PhysRev.105.1415}. The results clearly established that the muon is a structureless spin 1/2 Dirac particle, further solidifying its role as a fundamental tool in the study of the weak interaction.

Thanks to its relatively long lifetime and limited number of decay modes, the muon is an exceptionally clean probe of the dynamics of new phenomena. Over the decades, brighter muon beams and more sophisticated experiments have enabled increasingly stringent tests of the SM. To this day, the muon lifetime provides the best determination of the Fermi constant. Spectroscopic studies of muonic atoms and muonium, a pure QED $\mu^+ e^-$ bound state, yield unrivaled measurements of many fundamental quantities, such as the proton charge radius and the electron-to-muon mass ratio. The determination of the muon anomalous magnetic moment offers precise tests of the SM and places strong constraints on many speculative theories. Impressive limits on charged lepton flavor violating transitions are probing New Physics (NP) mass scales well above those directly accessible at colliders and are closely linked to the physics of neutrino masses, opening another portal into grand unified theory (GUT)-scale physics.

A global program is underway to further enhance the discovery potential of a wide range of muon probes, with significant increases in sensitivity anticipated over the next decade. In turn, these perspectives have catalyzed a renewed interest in models of light new particles and unknown forces accessible in muon reactions. In this review, we discuss recent experimental and theoretical advances in precision muon physics. Section~\ref{Sec:Decays} examines the status of muon decay parameters. The muon magnetic and electric dipole moments are covered in Section~\ref{Sec:Moments}, together with the current theoretical situation. Searches for charged lepton flavor violating transitions are finally surveyed in Section~\ref{Sec:CLFV} with an emphasis on the signature of light NP models. We refer the reader to several recent reviews for a more comprehensive discussion of these topics~\cite{Gorringe:2015cma, Keshavarzi:2022kpc, Calibbi:2017uvl,Bernstein:2013hba}.

\section{MUON DECAYS}
\label{Sec:Decays}
Muon decay into an electron and a neutrino-antineutrino pair ($\mu^- \to e^- \bar\nu_e \nu_\mu$), which is sometimes referred to as the ''Michel decay,'' is  one of the best-studied processes in particle physics. Accurately described by the Standard Model, it involves no strongly interacting particles in its initial or final states. Higher-order SM contributions, like QED or $1/m_W$ corrections, can be calculated with the required precision. In the SM at next-to-leading order in QED coupling corrections, the muon width can be expressed as 
\begin{equation}
\label{MuonWidthSM}
\Gamma_\mu^{\rm SM} = \frac{G_F^2 m_\mu^5}{192 \pi^3} F\left(\frac{m_e^2}{m_\mu^2}\right)
\left(1+\frac{3}{5} \frac{m_\mu^2}{m_W^2} \right)
\left[ 1+\frac{\alpha(m_\mu)}{2\pi} \left(\frac{25}{4} - \pi^2\right)\right]
\end{equation}
where $F(x) = 1 - 8x + 8 x^3 - x^4 - 12 x^2 \log x$. Following \cite{Kinoshita:1958ru,Marciano:1988vm}, both one-loop electromagnetic corrections and $1/m_W^2$ effects coming from the next-to-leading corrections to the Fermi model are included. 

The numerically leading corrections from heavy beyond the Standard Model (BSM) particles are expected to arise from dimension-six four-fermion operators. The corresponding normalized Lagrangian \cite{Petrov:2021idw}, valid below the EWSB scale, takes the form
\begin{eqnarray}
{\cal L}_{\ell_1 \to \ell_2 \nu_2 \bar\nu_1} &=& -\frac{4 G_F}{\sqrt{2}} \Big[
g_{RR}^S \left(\overline {\ell_2}_R {\nu_{\ell_2}}_L\right) \left(\overline {\nu_{\ell_1}}_L  {\ell_1}_R\right)
+ g_{RL}^S \left(\overline {\ell_2}_R {\nu_{\ell_2}}_L \right) \left(\overline {\nu_{\ell_1}}_R  {\ell_1}_L\right)
\nonumber \\
&+& \
g_{LR}^S \left(\overline {\ell_2}_L {\nu_{\ell_2}}_R \right) \left(\overline {\nu_{\ell_1}}_L  {\ell_1}_R\right)
+ g_{LL}^S \left(\overline {\ell_2}_L {\nu_{\ell_2}}_R \right) \left(\overline {\nu_{\ell_1}}_R  {\ell_1}_L\right)
\nonumber \\
&+& \
g_{RR}^V \left(\overline {\ell_2}_R \gamma^\alpha {\nu_{\ell_2}}_R\right) \left(\overline {\nu_{\ell_1}}_R \gamma_\alpha {\ell_1}_R\right)
+ g_{RL}^V \left(\overline {\ell_2}_R \gamma^\alpha {\nu_{\ell_2}}_R\right) \left(\overline {\nu_{\ell_1}}_L \gamma_\alpha {\ell_1}_L\right)
\\
&+& \
g_{LR}^V \left(\overline {\ell_2}_L \gamma^\alpha {\nu_{\ell_2}}_L\right) \left(\overline {\nu_{\ell_1}}_R \gamma_\alpha {\ell_1}_R\right)
+ g_{LL}^V \left(\overline {\ell_2}_R \gamma^\alpha {\nu_{\ell_2}}_R\right) \left(\overline {\nu_{\ell_1}}_L \gamma_\alpha {\ell_1}_L\right)
\nonumber \\
&+& 
\frac{g_{RL}^T}{2} \left(\overline {\ell_2}_R \sigma_{\alpha\beta} {\nu_{\ell_2}}_L \right) 
\left(\overline {\nu_{\ell_1}}_R  \sigma^{\alpha\beta} {\ell_1}_L\right)
+ \frac{g_{LR}^T}{2} \left(\overline {\ell_2}_L \sigma_{\alpha\beta} {\nu_{\ell_2}}_R \right) 
\left(\overline {\nu_{\ell_1}}_L  \sigma^{\alpha\beta} {\ell_1}_R\right)
+ h.c. ~ \Big]. \nonumber
\label{L_MeEGam}
\end{eqnarray}
%
%
%
This Lagrangian parametrizes the NP corrections to the dominant decay channel, $\mu\to e \nu \bar \nu$. The operator basis used here assumes the existence of right-handed neutrinos, and neglects flavor violation, which will be discussed in Section \ref{Sec:CLFV}. In the SM, all coupling constants are zero except for $g_{LL}^V=1$. 

\subsection{Michel parameters}
Michel parameters offer a model-independent framework for describing the energy and angular distributions in charged lepton decays. Originally introduced to characterize muon decays, the framework remains a powerful tool for probing the chiral structure of weak interactions and searching for deviations from the SM. The most general form of the differential decay rate for polarized muons is given by \cite{Kuno:1999jp}
\begin{equation}\label{MuonDecayNP}
\frac{d^2 \Gamma (\mu^\pm \to e^\pm \nu \bar\nu)}{d \cos\theta_e dx} = \frac{G_F^2 m_\mu}{4 \pi^3} W^4_{e\mu}
\sqrt{x^2-x_0^2} \left(F_{IS}(x) \pm P_{(\mu)} F_{AS}(x) \cos\theta_e \right) 
\left(1 + \vec P_e\cdot \hat{\vec\zeta} \right)
\end{equation}
where $W_{e\mu} = (m_\mu^2+m_e^2)/(2 m_\mu)$, $x=E_e/W_{e\mu}$, $x_0 = m_e/W_{e \mu}$, $\vec P_e$ is the polarization vector of the electron, and $\hat{\vec\zeta}$ is a unit vector along the direction of the measurement of electron (positron) spin polarization. The functions $F_{IS}(x)$ and $F_{AS}(x)$ are defined as
\begin{eqnarray}
F_{IS}(x) &=& x(1-x) + \frac{2\rho}{9} \left(4x^2-3x-x_0^2\right) + \eta x_0 (1-x)
\nonumber \\
F_{AS}(x) &=& \frac{\xi}{3} \sqrt{x^2-x_0^2} \left[
1-x+ \frac{2\delta}{3} \left(4x-3+\left(\sqrt{1-x_0^2}-1\right)\right)
\right]
\end{eqnarray}
where $\rho$, $\eta$, $\xi$, and $\delta$ are the so-called Michel parameters~\cite{Bouchiat:1957zz}. In the Standard Model $\rho=\delta=3/4$, $\eta = 0$, and $\xi =1$. The description of the Michel parameters in terms of the coefficients $g_{XX}^L$ of the effective Lagrangian of Eq.~(\ref{L_MeEGam}) can be found in ~\cite{Petrov:2021idw}. 
The muon decay width, including Michel parameters, can be written as 
\begin{eqnarray}\label{MuonWidthNP}
\Gamma_\mu &=& \frac{G_F^2 m_\mu^5}{192 \pi^3} \left[
F\left(\frac{m_e^2}{m_\mu^2}\right) + 4 \eta \frac{m_e}{m_\mu} G\left(\frac{m_e^2}{m_\mu^2}\right)
- \frac{32}{3} \frac{m_e^2}{m_\mu^2} \left(\rho-\frac{3}{4}\right) \left(1-\frac{m_e^4}{m_\mu^4}\right)
\right]
\nonumber \\
&\times& \left(1+\frac{3}{5} \frac{m_\mu^2}{m_W^2} \right)
\left[ 1+\frac{\alpha(m_\mu)}{2\pi} \left(\frac{25}{4} - \pi^2\right)\right]
\end{eqnarray}
where $G(x) = 1 + 9x - 9 x^2 - x^3 + 6x(1+x) x^2 \log x$. It is worth noting that Eq.~(\ref{MuonWidthNP}) describes the decay $\mu \to e + \slashed{E}$, where $\slashed{E}$ denotes missing energy, and assumes that no new decay channels from light NP particles coupling to the muon are present. 

Experimentally, the most precise determination of the Michel parameters was performed by the TRIUMF Weak Interaction Symmetry Test (TWIST) experiment~\cite{TWIST:2011aa, TWIST:2011egd} by analyzing the energy and angular distributions of positrons emitted in the decay of highly polarized muons. A muon beam was first produced by the decays of pions created in the interactions of high-energy protons in a graphite target. A fraction of these pions stopped near the target surface, yielding a nearly 100\% polarized, mono-energetic muon beam. These surface muons were then transported without depolarization towards the experimental area and stopped in a thin target foil at the center of the detector. The latter comprised a series of planar drift chambers and proportional chambers immersed in a 2 T magnetic field to precisely measure the trajectory of decay positrons. The Michel decay parameters extracted from an analysis of about ten billion muon decays were as follows (the parameter $\xi$ and the muon polarization in pion decays $P^\pi_\mu$ are experimentally inseparable):
\begin{eqnarray*}
\rho &=& 0.74977 \pm 0.00012 \rm \, (stat) \pm 0.00023 \, (syst) \\
\delta &=& 0.75049 \pm 0.00021 \rm \, (stat) \pm 0.00073 \, (syst)\\
P^\pi_\mu \xi &=& 1.00084 \pm 0:00029 \rm \, (stat) \pm ^{0.00065}_{0.00063} \, (syst)\\
\end{eqnarray*}
consistent with the SM expectations at the level of $\sim 10^{-3}$. A global analysis of decay parameters in terms of the coupling constants $g_{XY}^Z$ can be found in~\cite{ParticleDataGroup:2024cfk}, and constraints on left-right symmetric extensions of the SM electroweak interaction~\cite{Herczeg:1985cx} are explored, for example, in \cite{TWIST:2011egd}.

\section{MAGNETIC AND ELECTRIC DIPOLE MOMENTS}
\label{Sec:Moments}
\subsection{Theoretical considerations}
A charged spin-$1/2$ particle has a magnetic dipole moment that can be expressed as
\beq
\label{MagMomAbs}
\vec \mu=g \frac{e}{2 m} \vec s
\eeq
where $\vec\mu$ is the spin magnetic moment of the particle, $m$ is its mass, and $e$ is the elementary charge. For a structureless, point particle, the $g$-factor is $g=2$. In non-relativistic quantum mechanics, the interactions between a Dirac particle's spin and an external magnetic field can be described by Pauli's Hamiltonian.
\beq\label{MagMom}
{\cal H}_M = - \vec \mu \cdot \vec B = - \mu \ \vec \sigma \cdot \vec B
\eeq
where $\sigma_i$ are the Pauli matrices, and $\vec\mu$ is the magnetic dipole moment defined in Eq.~(\ref{MagMomAbs}). The latter must be aligned with the spin direction, as that is the only available orientation for a spin-1/2 particle.

\begin{figure}[ht]
    \centering   
    \includegraphics[width=0.8\textwidth]{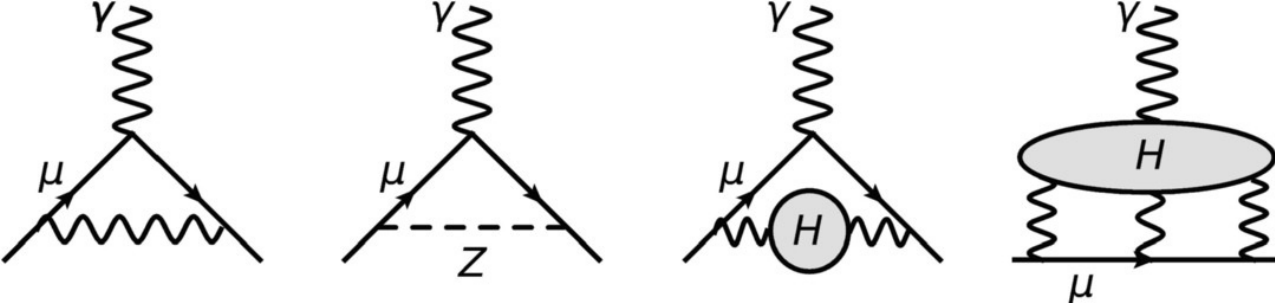}
    \caption{The vacuum polarization contribution to the anomalous magnetic moment of the muon. The diagrams show (from left to right) the one-loop QED correction, the one-loop Z-boson exchange, the leading-order hadronic vacuum polarization diagram, and the hadronic light-by-light contribution. Figures adapted from References~\cite{Keshavarzi:2022kpc} (CC-BY).}
    \label{fig:AnomMagMom}
\end{figure}

Quantum corrections modify the value of the $g$-factor. The probe, a photon, can interact with one of the virtual particles responsible for quantum corrections to the photon-lepton vertex, rather than the lepton itself (see Figure~\ref{fig:AnomMagMom}). This allows exploration of the "quantum structure" of the vertex. It is helpful to define the magnetic moment in terms of general form factors used in lepton-photon interactions. The most general matrix element of the electromagnetic current $j^\mu$ can be written as.
\beq
\langle p_i | j^\mu |p_f \rangle = 
\overline{u} (p_i) \Bigl[\Gamma^\mu_{\CPc} (P,q) + \Gamma^\mu_{\CPv} (P,q) \Bigr] u(p_f)
\label{eq:Gcp}
\eeq
where the generalized vertex $\Gamma^\mu_{\CPc}$ describes the $\CP$ conserving part of the interaction, while $\Gamma^\mu_{\CPv}$ contains the part that might break $\CP$ symmetry. The $\CP$-conserving part is
\beq\label{Formfactors1}
\Gamma^\mu_{\CPc} (P,q) = (-ie) \ \left[
F_1(q^2) \gamma^\mu + \frac{iF_2(q^2)}{2m} \sigma^{\mu\nu} q_\nu 
\right]
\eeq
where $q=p_f-p_i$, $P=p_f+p_i$, and $u(p_i)$ and $u(p_f)$ are the on-shell 4-spinors normalized such that $\overline{u}(p) u(p) = 2 m$. The magnetic moment can then be defined as $\mu=(F_1(0)+F_2(0))/(2 m)$, and $g=F_1(0)+F_2(0)$ in Eq.~(\ref{MagMomAbs}). 

Although quantum corrections modify the electromagnetic vertex in Eq.~(\ref{Formfactors1}), gauge invariance ensures that $F_1(0)=1$ remains unchanged, i.e., $F_1(0)$ is not renormalized by QED corrections. By contrast, $F_2(0)$ is renormalized, resulting in a deviation of the $g$-factor from its non-renormalized value of two. The {\it anomalous magnetic moment} is defined as a measurement of this deviation:
\beq
a_f = \frac{g-2}{2} = F_2(0)
\eeq
where the index $f$ indicates the lepton flavor. 

The anomalous magnetic moment can be calculated within the Standard Model with excellent precision. It can, therefore, be used to compare with the experimentally measured value to identify potential contributions from New Physics. The Standard Model prediction for the muon anomalous magnetic moment can be expressed as a sum of three contributions,
\beq\label{Breakdown}
a_\mu^{SM} = a_\mu^{QED} + a_\mu^{EW} + a_\mu^{had}.
\eeq
The leading order QED contribution $a_\mu^{QED,LO} = \alpha/(2\pi)$ was one of the first radiative corrections computed. It is known as the Schwinger term and happens to be finite \cite{Schwinger:1948iu}.


The QED contribution, which includes all the photonic and leptonic loops, and the electroweak contribution, containing the loops involving $Z$, $W$, or Higgs bosons, are known very accurately \cite{Aoyama:2012wk},
 \bea\label{Gm2QED}
 a_\mu^{QED} &=&  116 584 718.951 (0.009)(0.019)(0.007)(.077) \times 10^{-11} \nonumber \\
 a_\mu^{EW}  &=& 153.6(1.0) \times 10^{-11}.
 \eea
The numerical results shown in Eq.~(\ref{Gm2QED}) include QED corrections calculated up to five loops, and electroweak corrections calculated to two loops, with the leading three-loop contributions also included.

The primary uncertainty in the theoretical prediction of $a_\mu$ is in the final term of Eq.~(\ref{Breakdown}), which encompasses all hadronic loop contributions (Fig.~\ref{fig:AnomMagMom}). This term can also be divided into two main parts.
\beq\label{HadrBreakdown}
a_\mu^{had} = a_\mu^{hvp} + a_\mu^{hlbl}
\eeq
where $a_\mu^{hvp}$ represents the hadronic vacuum polarization (HVP) contribution, and $a_\mu^{hlbl}$ denotes the hadronic light-by-light scattering contribution. The two-loop hadronic vacuum polarization involves integrating over two momenta, including the one flowing through the vacuum polarization loop. For quarks running inside the loop, integration over all distances is necessary to obtain a complete hadronic contribution. While this is not an issue for similar loops with leptons, which are calculated in the QED and electroweak contributions to $a_\mu$, the quark contributions are more delicate to evaluate. This is because perturbative QCD may not reliably estimate the total quark contribution, especially at low momenta. In that region, more sophisticated methods are necessary to accurately account for the hadronic vacuum polarization \cite{Aliberti:2025beg}.

Two main methods are used to calculate these contributions. The more phenomenological approach involves using experimental data from $e^+ e^-$ annihilation (or hadronic $\tau$-lepton decay) to evaluate the vacuum polarization contribution from hadrons. The other method relies on lattice QCD techniques and involves directly calculating $\Pi(Q^2)$ in discretized space-time. While those approaches are quite different, they both include the evaluation of the vacuum polarization loop $\Pi(q^2)$, defined from the tensor
\beq\label{VacPol}
\Pi_{\mu\nu}(q) = i \int d^4 x e^{i x\cdot q} 
\langle 0| T\left[J_\mu(x) J_\nu(0) \right] | 0 \rangle = 
\left(q_\mu q_\nu - q^2 g_{\mu\nu} \right) \Pi(Q^2)
\eeq
where $Q^2 = -q^2$. The current $J_\mu (x)$ denotes the electromagnetic current of the light quarks,
\beq\label{QCurrent}
J_\mu (x) = \sum_q Q_q \ {\overline q}(x) \gamma_\mu q(x)
\eeq
for the quark flavor $q=u,d,s$ with $Q_u=2/3$ and $Q_d = Q_s = -1/3$. 
It follows from analyticity of vacuum polarization function that $\Pi(q^2)$ satisfies a once-subtracted dispersion relation \cite{Peskin:1995ev},
\beq\label{DispRel2}
\Pi(q^2) = \frac{q^2}{\pi}  \int_{s_0}^\infty \frac{ds}{s(s-q^2)} \mbox{Im}\Pi(s)
\eeq
where $\Pi(0) = 0$, as required by derivative couplings of pions in the chiral limit.  The key observation is that the dispersive part of $\Pi(s)$ can be related to 
a total cross section for $e^+ e^-$ annihilation into hadrons,
\beq
\mbox{Im}\Pi(s) = \frac{1}{12 \pi} R(s) = \frac{1}{12 \pi} \frac{\sigma(e^+e^- \to \rm hadrons)}
{\sigma(e^+ e^- \to \mu^+\mu^-)}
\eeq
where $R(s)$ represents the experimentally measured $R$-ratio of the cross section for $e^+e^-\to$ hadrons. The HVP can be written as a convolution
\beq\label{HVP}
a_\mu^{hvp} = \frac{1}{3} \left(\frac{\alpha}{\pi}\right)^2 \int_{4 m_\pi^2}^\infty
\frac{ds}{s} K(s) R(s)
\eeq
with a QED kernel $K(s)$ given by
\beq\label{Kernel}
K(s) = \int_0^1 dx \frac{x^2(1-x)m_\mu^2}{x^2 m_\mu^2+(1-x)s}.
\eeq
As seen from Eqs.~(\ref{HVP}) and (\ref{Kernel}), the expression for $a_\mu^{hvp}$ is very sensitive to the low-energy part. The uncertainty in evaluating Eq.~(\ref{HVP}) comes, therefore, from the contribution of the two-pion state, which is currently limited by a tension between the experimental data sets from the KLOE, \babar, and CMD-3 experiments. The same data could be obtained from the hadronic decays of $\tau$-leptons, although controlling the isospin-violating corrections has proven difficult \cite{Aliberti:2025beg}. 

The lattice QCD calculation of the hadronic vacuum polarization contribution from first principles is also possible. Recently, a precise result for $a_\mu^{hvp}$ has been obtained through the direct computation of Eq.~(\ref{VacPol}). In particular, the HVP contribution on the lattice can be written as
\beq\label{HVPLattice}
a_\mu^{hvp,l} = \left(\frac{\alpha}{\pi}\right)^2 \int_0^\infty ds G(s) \widetilde K(s)
\eeq
with 
\beq
G(s) = -\frac{a^3}{3} \sum_{k=1}^3 \sum_{\vec x} \langle 0| J_k (x) J_k(0) | 0 \rangle
\eeq
where $\widetilde K(s)$ is a known analytic QED kernel function, which, just like the $K(s)$ in Eq.~(\ref{Kernel}), gives weight to the long-distance regime of the correlation function, and $a$ is the discretization length \cite{Aliberti:2025beg}.

Since anomalous magnetic moments of leptons arise from radiative corrections, they can be affected by potential contributions from virtual NP particles. As there are many models, the SMEFT Lagrangian offers a convenient way to parameterize them in a single framework. The relevant Lagrangian is~\cite{Fortuna:2024rqp}:
\bea\label{MM_SMEFT}
{\cal L}_{\rm (g-2)} &=&
\frac{C^\ell_{eB}}{\Lambda^2}  \ \overline L_L \sigma^{\mu\nu} H \ell_R B_{\mu\nu}
+ \frac{C^\ell_{eW}}{\Lambda^2} \ \overline L_L \sigma^{\mu\nu} \tau^I H \ell_R W^I_{\mu\nu} + h.c.
\nonumber \\
&= & \frac{C^\ell_{e\gamma}}{\Lambda^2} \frac{v}{\sqrt{2}} \overline \ell_L \sigma^{\mu\nu} \ell_R F_{\mu\nu}
+ \frac{C^\ell_{eZ}}{\Lambda^2} \frac{v}{\sqrt{2}} \overline L_f \sigma^{\mu\nu} \ell_R Z_{\mu\nu} + h.c.,
\eea
where $\Lambda$ is the New Physics mass scale, $L_L$ denotes the left-handed lepton doublet, and $H$ represents the Higgs doublet fields. The $W^I_{\mu\nu}$ and $B_{\mu\nu}$ are the weak and hypercharge fields of the Standard Model before the spontaneous symmetry breaking (SSB). The second line of Eq.~(\ref{MM_SMEFT}) is derived after the SSB. 
The NP contribution to $\delta a_\mu$ is given by \cite{Fortuna:2024rqp}:
\beq
\delta a_\mu = \frac{4 m_\mu v}{\sqrt{2} e \Lambda^2} \Bigl[C^\mu_{e\gamma} -
C^\mu_{eZ}  \frac{3 \alpha}{2\pi} \cot(2\theta_{\rm W}) \log \frac{\Lambda^2}{m_Z^2} \Bigr]
\eeq
where $\theta_{\rm W}$ is the weak mixing angle. 

The $\Gamma^\mu_{\CPv}$ part of the vertex in Eq.~\ref{eq:Gcp} can be used to define the {\it electric dipole moment}. It can be written as
\beq\label{FormfactorsEDM}
\Gamma^\mu_{\CPv} (P,q) = (-ie) \ \left[ F_A(q^2) \left(\gamma^\mu \gamma_5 q^2 - 2m \gamma_5 q^\mu\right)
+ \frac{F_3(q^2)}{2m} \sigma^{\mu\nu} \gamma_5 q_\nu \right].
\eeq
The electric dipole moment can now be expressed as 
\beq\label{FormfactorsEDM3}
d_f = -\frac{F_3(0)}{2 m}.
\eeq
Similar to Eq.~(\ref{MagMom}), the effective Hamiltonian generating an electric dipole moment (EDM) can be defined as
\beq\label{EDM}
{\cal H}_E = - \vec d \cdot \vec E = - d \ \vec \sigma \cdot \vec E.
\eeq
This Hamiltonian describes the interaction of a particle's dipole moment $\vec d$ with an electric field $\vec E$. As with the magnetic moment, it must align with the spin. Unlike the magnetic moment, ${\cal H}_E$ is {\it not invariant} under a combined $\CP$ transformation \cite{Petrov:2021idw}. This difference is crucial: the electric dipole moment is only nonzero when $\CP$-violating interactions exist. In the minimal Standard Model, contributions to $\CP$-violating amplitudes originate from the quark sector and are proportional to the Jarlskog invariant $J$~\cite{Jarlskog:1985ht}. It can be shown that the non-zero SM contribution to $d$ first appears at four-loop order \cite{Petrov:2021idw}, implying that only CP-violating parts of potential new physics interactions can be tested by examining the EDM of various particles. In many specific NP models with other CP-violating mechanisms, EDMs can be produced at lower loop levels than in the SM, making them very sensitive probes.

It is straightforward to derive the expression for the EDM via a tree-level matching with the SMEFT Lagrangian. Two operators are serving this purpose:
\bea\label{EDM_eff}
{\cal L}_{\rm EDM} =  \frac{C_{eW}}{\Lambda^2} \overline L_f \sigma^{\mu\nu} \tau^I H \ell_f W^I_{\mu\nu}
+  \frac{C_{eB}}{\Lambda^2}  \overline L_f \sigma^{\mu\nu} H \ell_f B_{\mu\nu} .
\eea
In the broken phase, the Higgs field in Eq.~(\ref{EDM_eff}) acquires a vacuum expectation value $v$, just like in Eq.~(\ref{MM_SMEFT}). Separating out the $W_\mu^0$ and $B_\mu$ contributions, rotating to the physical basis of $Z_\mu$ and a photon $A_\mu$, and taking the imaginary part of the coefficient gives 
\beq\label{EDM_SMEFT}
d_\ell (\mu) = \frac{\sqrt{2} v}{\Lambda^2} \mbox{Im}
\left[s_{\rm W} C_{\ell W} (\mu) - c_{\rm W} C_{\ell B} (\mu) \right],
\eeq
where $s_{\rm W}$ and $c_{\rm W}$ are the sine and cosine of the Weinberg angle. It should be noted that EDM measurements could also be sensitive to Wilson coefficients other than $C_{\ell W}$ and $C_{eB}$ since other operators might mix with those in Eq.~(\ref{EDM_eff}) at one or more loop levels.

It is sometimes helpful to write an effective Lagrangian whose Wilson coefficients are proportional to $a_f$ and $d_f$ for different lepton flavors $f$. Such an effective Lagrangian can be written as \cite{Barbieri:1974nc,Jegerlehner:2009ry}
\beq\label{EffLagAMM}
{\cal L}_{\rm eff} = - \frac{1}{2} \overline \psi_{\ell_f} \sigma^{\mu\nu} 
\left[ D_{\ell_f} P_R + D_{\ell_f}^* P_L\right] \psi_{\ell_f} F_{\mu\nu}
\eeq
where $\psi_{\ell_f}$ represents a lepton field of flavor $f$. One can see that 
\beq\label{EffLagAMM_WC}
\mbox{Re} \ D_{\ell_f} = \frac{a_f e}{2 m_f}, \qquad 
\mbox{Im} \ D_{\ell_f} = \frac{\eta_f}{2} \frac{e}{2 m_f} = d_f,
\eeq
\ie, the imaginary part of the $D_{\ell_f}$ (and thus $F_2(0)$) corresponds to an EDM of a lepton.

\subsection{Magnetic dipole moment measurement}
The determination of the muon anomalous magnetic moment has progressed over several decades, marked by a steady progression in experimental and theoretical precision. The first measurements, notably those performed at the Nevis cyclotron of Columbia University in 1957~\cite{PhysRev.105.1415}, analyzed the Larmor precession of stopped muons in a magnetic field to determine the corresponding $g$-factor. 
While the results corroborated the Dirac predictions, the precision remained insufficient to measure the anomalous component. A leap in sensitivity was achieved through a series of pioneering experiments at CERN with the introduction of muon storage rings. By injecting a polarized muon beam and measuring its spin evolution, the anomalous magnetic moment could be directly determined from the difference between the cyclotron frequency and the muon spin precession frequency. The combination of all CERN measurements reached an impressive precision of 7.3 ppm~\cite{CERN-Mainz-Daresbury:1978ccd}, consistent with the increasingly sophisticated QED calculations. This approach evolved with the E821 experiment at Brookhaven National Laboratory, introducing superconducting magnets and a novel muon injection technique to reduce the uncertainty down to 0.54 ppm~\cite{Muong-2:2006rrc}. An intriguing discrepancy with the SM predictions motivated the pursuit of this quest with the Muon $g$-2 experiment at Fermilab~\cite{Muong-2:2015xgu}.

\begin{figure}[ht]
    \includegraphics[width=0.6\textwidth]{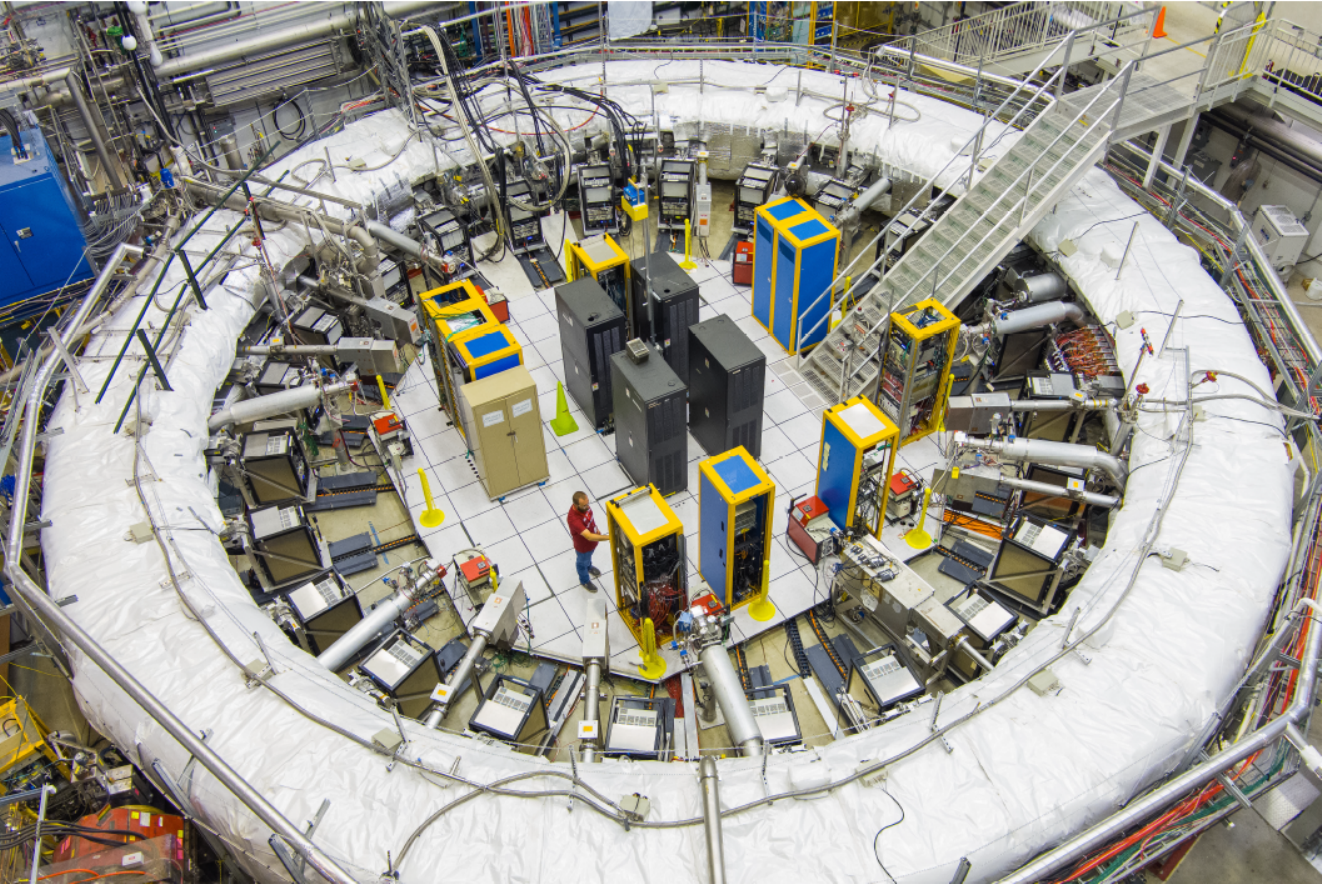}
    \caption{view of the muon g-2 experiment at Fermilab. Three superconducting coils spanning the circumference of the steel yoke provide a 1.45 T magnetic field, and four electrostatic quadrupoles positioned symmetrically around the ring provide vertical beam confinement. The storage ring is covered by a white insulating blanket. Figure reproduced from https://muon-g-2.fnal.gov with permission from Fermilab.}
    \label{fig:g2ring}
\end{figure}

Similar to its predecessor, the Muon $g$-2 experiment measures $a_\mu$ by injecting polarized muons into a storage ring and observing their spin dynamics through their decays. The 7.1 m radius ring used in the previous BNL experiment was relocated on the Fermilab muon campus and upgraded to increase the field uniformity (see Figure~\ref{fig:g2ring}). In the presence of a constant magnetic field, muons circulate the ring with a cyclotron frequency $\omega_c$ while their spins rotate with a frequency $\omega_S$ proportional to the muon $g$ factor. The frequency difference, called the anomalous spin precession frequency $\omega_a$, is (nearly) proportional to the magnetic anomaly
\begin{equation}
    \vec{\omega}_a = \vec{\omega_c} - \vec{\omega_s} = -\frac{q}{m_\mu} \left[  a_\mu \vec{B} -a_\mu \left( \frac{\gamma}{\gamma+1}\right)  (\vec{\beta} \cdot \vec{B})\vec{B}  - \left(a_\mu - \frac{1}{\gamma^2-1}\right) \frac{\vec{\beta} \times \vec{E}}{c}  \right]
\end{equation}
where $\vec{E}$ and $\vec{B}$ are the electric and magnetic fields present in the ring and $\vec{\beta}$ is the muon velocity. For muons injected at the "magic momentum" of $3.09 \GeVc$ in an orbit perpendicular to the magnetic field, both the second and the third terms vanish, and this equation reduces to $\omega_a = \frac{a_\mu q}{m_\mu}B$.  

The magnetic anomaly is encoded in the energy distribution of the positrons emitted by the muon decays as they orbit the ring. Due to parity violation, the high-momentum positrons are preferentially emitted along the muon spin direction. In the laboratory frame, the energy distribution depends on the relative angle between the muon spin and momentum; the energy spectrum is the hardest when they are aligned. The positrons are recorded by a set of calorimeters positioned around the inner circumference of the ring. The number of high-energy positrons over time $N(t)$ is therefore modulated by the precession frequency, and the magnitude of $\vec{\omega}_a$ can be extracted by a fit of the form $N(t) = N_0 e^{-t/\gamma \tau} \left[1 + A \cos(\omega_a t + \phi_0) \right]$, where $\tau$ is the muon lifetime, $A$ is the muon asymmetry, and $\phi_0$ is an initial phase. A series of corrections is applied to account for beam dynamics effects due to the electric field, the beam momentum spread, lost muons, and the correlation between a muon initial phase and its probability of being detected. 

The magnetic field is measured with nuclear magnetic resonance (NMR) probes mounted on a trolley circulated inside the ring. Variations in the magnetic field between trolley runs are continuously monitored with 378 NMR probes located around the ring. The resulting maps are weighted by the spatial and temporal distributions of the muons determined by two straw tube tracking detectors installed around the ring. Corrections are applied to account for eddy currents in the kicker magnet during beam injection and perturbations from vibrations of the beam focusing system. 

\begin{figure}[ht!]
    \centering   
    \includegraphics[width=0.7\textwidth]{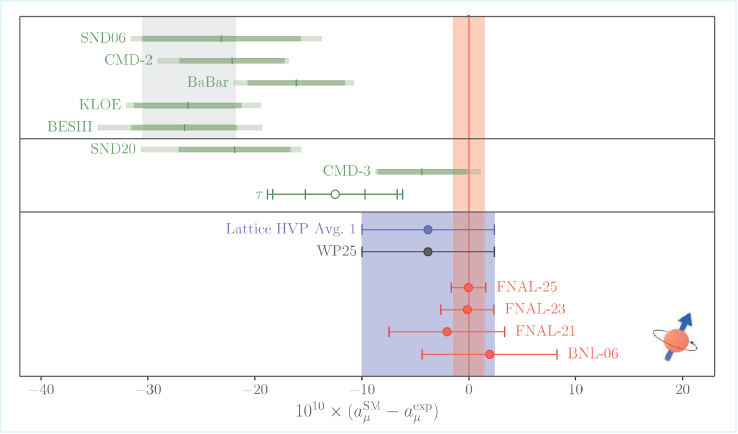}
    \caption{The experimental values of the muon anomalous magnetic moment (red band) together with theoretical predictions based on lattice QCD calculations (blue band) and data-driven methods (gray band). Figure adapted from Reference~\cite{Aliberti:2025beg} (CC-BY).}
    \label{fig:gm2}
\end{figure}

By expressing the magnetic field in terms of the proton Larmor frequency $\omega_p$, the anomalous magnetic moment can be written as a product of precisely measured quantities:
\begin{equation}
a_\mu = \frac{\omega_a}{\omega_p}\frac{\mu_p}{\mu_e}\frac{m_\mu}{m_e}\frac{g_e}{2}
\end{equation}
where $\mu_p/\mu_e$ is the ratio of the proton-to-electron magnetic moments and $g_e$ is the electron $g$-factor. Data were collected over six campaigns starting in 2018, increasing the statistics of previous experiments by a factor of 2.5. Over time, several upgrades were implemented to minimize systematic uncertainties, particularly those arising from the electrostatic quadrupole system, the kicker magnet, and coherent betatron oscillations. The final result, displayed in Figure~\ref{fig:gm2}, reaches a precision of 127 ppb~\cite{Muong-2:2025xyk}, a four-fold improvement over the previous BNL measurement. The results agree with the SM expectations based on lattice QCD calculations, while a $\sim 5\sigma$ discrepancy is observed with the theoretical value estimated from data-driven measurements. Using the 2025 Theory Initiative SM update, a single SMEFT photon-dipole explaining the central shift ($\delta a_\mu = 38(63)\times 10^{-11}$) would correspond to probing an effective NP scale of about $800 \TeVcc$ if the Wilson coefficient is ${\cal O}(1)$ for tree-level-like NP, or $60-70 \TeVcc$ for loop-suppressed NP. Additional efforts are underway to understand the source of the deviation between the lattice and data-driven predictions. Notably, the MUonE experiment~\cite{MUonE:2016hru} aims to provide an independent measurement of the leading-order hadronic contribution to the muon anomalous magnetic moment by analyzing elastic muon scattering off atomic electrons in fixed targets. 

The muon $g$-2/EDM collaboration at J-PARC (E34)~\cite{Abe:2019thb} aims to perform a measurement of the muon $g$-2 and its electric dipole moment with an alternative approach. The key feature is the usage of a low-emittance muon beam generated by re-accelerating thermal muons. This technique eliminates the need for an electric field focusing, allowing the muon momentum to differ from the magic momentum without any significant disadvantage. A compact 3 T superconducting magnet generates a highly uniform magnetic field, confining the muons with a muon cyclotron radius of $\sim 33$ cm - about twenty times smaller than the Fermilab ring. The experiment targets a precision of 450 ppb with a systematic uncertainty of 70 ppb or less.

\subsection{Electric dipole moment measurement}
In the presence of an electric and magnetic field, an electric dipole moment (EDM) introduces an additional term to the muon precession $\omega$:
\begin{equation} \vec{\omega} = \vec{\omega}_a + \vec{\omega}_{EDM} = \vec{\omega}_a -\frac{\eta e}{2 m_\mu c} \left[ \vec{E} + c \vec{\beta} \times \vec{B} \right]
\end{equation}
where $\omega_a$ is the anomalous precession frequency and $\eta$ is a dimensionless factor describing the strength of the EDM. In a typical storage ring configuration - where the muons are circulating in a plane perpendicular to the magnetic field - the EDM-induced precession $\vec{\omega}_{EDM}$ points radially, orthogonal to $\omega_a$. As a result, the muon polarization plane is slightly tilted with respect to the ring plane, inducing an oscillation of the vertical decay positron angle as a function of time. An EDM also increases the magnitude of the precession frequency, but its effect is marginal given the current experimental constraints. Based on the analysis of the $g$-2 precession plane, the BNL $g$-2 experiment set a limit $d_\mu < 1.9 \times 10^{-19} \rm \, e \cdot cm$ (95\% CL)~\cite{Muong-2:2008ebm}. Improvements in sensitivity down to the level of $\sim 10^{-21} \rm \, e \cdot cm$ are expected at the Fermilab Muon $g$-2 and the J-PARC Muon $g$-2/EDM experiments.

A complementary strategy based on the frozen-spin technique is being pursued by the dedicated muEDM experiment at PSI~\cite{Sakurai:2022tbk}. By applying an electric field $E  \sim a_\mu B c \beta \gamma^2$, the contribution of $\omega_a$ to the precession frequency vanishes, and only the EDM term remains. In the absence of an EDM, the spin vector is effectively "frozen" in the direction of the momentum vector. The EDM signal is then observed as an up-down asymmetry of the positron count measured by detectors positioned around the ring. This experiment aims to reach a sensitivity at the level of $10^{-23} \rm \, e \cdot cm$.

\section{LEPTON FLAVOR VIOLATING INTERACTIONS}
\label{Sec:CLFV}
\subsection{Theoretical considerations}
\label{Sec:theoryLFV}
\begin{figure}[ht]
    \includegraphics[width=0.8\textwidth]{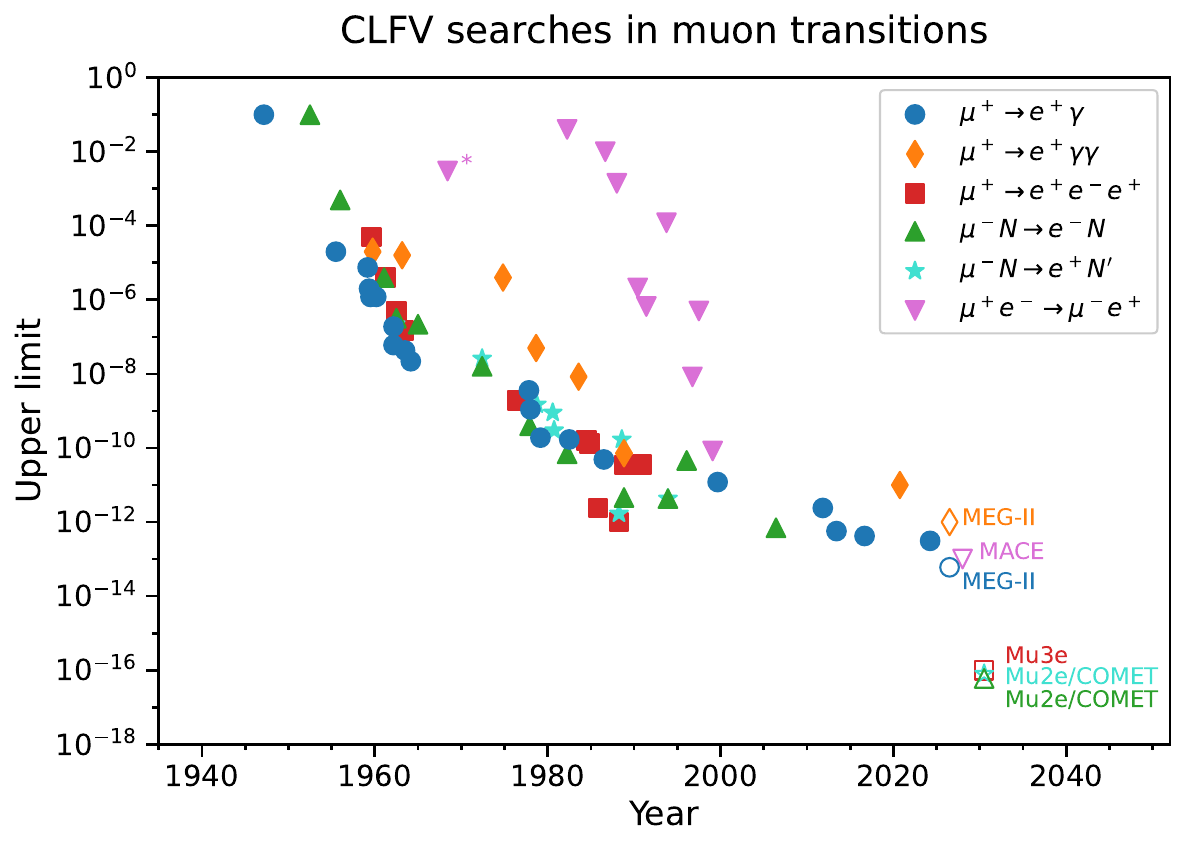}
    \caption{Historical evolution of CLFV searches in muons transitions. Experiments using stopped pion beams in the mid 1950s started to improve upon limits based on cosmic ray measurements, before stopped muon beams allowed further gain in the mid 1970s. The following limits are reported: the branching fractions of the $\mu^+ \rightarrow e^+ \gamma$, $\mu^+ \rightarrow e^+e^-e^+$, and $\mu^+ \rightarrow e^+ \gamma \gamma$ reactions, the ratio of the conversion rate to the total muon capture rate for the $\mu^- N \rightarrow e^- N$ and $\mu^- N \rightarrow e^+ N'$ reactions, and the muoniun-antimuonium conversion probability (the early measurement in Ar gas, indicated by an asterisk, also includes the probability of breaking the antimounium system and forming a muonic Ar atom). Solid symbols indicate past experiments, and open symbols denote the expected reach of future proposals. See References~\cite{Bernstein:2013hba, ParticleDataGroup:2024cfk} for details.}
    \label{fig:CLFVhistory}
\end{figure}

Lepton flavor violating decays are forbidden in the Standard Model at tree level. While these processes can occur in the SM at one loop level, their rates are extremely suppressed, as they are proportional to powers of neutrino masses. A common way to study lepton flavor violation is through processes involving only leptons (and possibly photons) in the initial and final states. Three muon transitions have consistently provided strong constraints over the past decades: the $\mu^+ \rightarrow e^+ \gamma$ and $\mu^+ \rightarrow e^+e^-e^+$ decays, and neutrinoless $\mu^- N \rightarrow e^- N$ conversion (see Figure~\ref{fig:CLFVhistory}). For heavy NP scenarios, these processes are governed by the following effective Lagrangian in the low-energy effective theory (LEET) with dimension five and six operators, 
\begin{eqnarray}\label{EffLagFCNC}
{\cal L} = \frac{1}{\Lambda^2_{\rm LFV}} \Bigl[
&& C_D \ \mu_\mu (\overline e \sigma^{\alpha\beta} P_R \mu) F_{\alpha\beta} 
+ C_S \ (\overline e P_R \mu) (\overline e P_R  e)
+ C_{VR} \ (\overline e \gamma^\alpha P_R \mu) 
(\overline e \gamma_\alpha P_R  e)
\nonumber \\
&+& C_{VL} \ (\overline e \gamma^\alpha P_L \mu)
(\overline e \gamma_\alpha P_L e)
+ \sum_{\Gamma,\Gamma'} C_{\Gamma \Gamma'} \ 
(\overline e \Gamma \mu) (\overline q \Gamma' q)
\Bigr]
\end{eqnarray}
where $\Lambda_{LFV}$ is the NP mass scale. The Wilson coefficients $C_i$ in Eq.~(\ref{EffLagFCNC}) can be viewed as components of a vector $\vec C$ normalized to one at the experimental scale. This Lagrangian, which accurately describes $\mu \to e \gamma$, $\mu \to 3 e$, and muon conversion at scales below spontaneous EWSB, can be matched to the SMEFT Lagrangian at higher scales and to the Lagrangian representing a specific NP model. In fact, the renormalization group equations can be used to relate the scales at $\Lambda_{LFV}$ and the experimental scale,
\beq
\vec C(m_\mu) = {\bf{G}}^T(\Lambda_{LFV},m_\mu) \vec C(\Lambda_{LFV})
\eeq
where the matrix $\bf{G}$ accounts for changing particle content at various scales \cite{Davidson:2022nnl}.

The simplest process to consider is a two-body radiative decay of a lepton state $\mu \to e \gamma$. The most general transition amplitude for $A_{\mu \to e \gamma}$ can be written by employing gauge invariance and equations of motion, 
\beq\label{MuEGamma}
A_{\mu \to e \gamma}(p,p^\prime) =\frac{i}{m_{\ell_1}} \overline u_{e}(p^\prime) 
\left[
A_L P_L + A_R P_R
\right] \sigma_{\alpha\beta} q^\beta
u_{\mu} (p) \epsilon^{*\alpha}
\eeq
where the projection operators are defined as $P_{L/R} = (1\mp\gamma_5)/2$, and $\epsilon^\alpha$ is the polarization vector of the photon. The projection operators can be applied to the electron spinor $u_{e}(p^\prime)$, defining the decay branching ratios into the left-handed ${\cal B}(\mu \to e_L \gamma)$ or the right-handed ${\cal B}(\mu \to e_R \gamma)$ electrons. The unpolarized decay rate of the muon is 
\beq
\Gamma(\mu \to e \gamma) = \frac{m_{\mu}}{16 \pi} \left(
\left| A_L\right|^2 + \left| A_R\right|^2
\right).
\eeq
In the SM, the rate $\Gamma(\mu \to e \gamma)$ is given by \cite{Bilenky:1977du}
\beq
{\cal B} (\mu \to e \gamma)
= \frac{3\alpha}{32 \pi} \left|\sum_i U_{\mu i}^* U_{ei} \frac{m_i^2}{m_W^2}\right|^2
\eeq
where $U_{ik}$ is the neutrino mixing matrix and $m_i$ are the corresponding masses of neutrino mass eigenstates. Numerically, this predicts the associated branching ratio in the SM to be ${\cal B}  (\mu \to e\gamma)_{\nu SM} \sim 10^{-54}$. Similar results can be obtained for the $\mu \to 3e$ and the muon conversion rates, making those decays SM background-free tests of New Physics \cite{Petrov:2021idw}.

It is interesting to note that the coefficients $A_L$ and $A_R$, in principle, depend on all the Wilson coefficients in the LEET Lagrangian (\ref{EffLagFCNC}) if higher-order loop corrections are considered. At leading order, the final results rely on the scalar products of the vector $\vec C$ and the unit vectors defining the decay in Eq.~(\ref{EffLagFCNC}). For example, for the $\mu \to e_L \gamma$ transition, the result can be expressed as \cite{Davidson:2022nnl}
\beq
{\cal B} (\mu \to e_L \gamma) = 384 \pi^2 \frac{v^4}{\Lambda_{\rm LFV}^4} \left| \vec C \cdot \hat e_{DR}\right|^2 < {\cal B}^{\rm exp}_{\mu \to e\gamma}
\eeq
where we, following~\cite{Davidson:2022nnl}, introduced unit vectors $\hat e_{DR}$ which selects the coefficient of the dipole operator in Eq.~(\ref{EffLagFCNC}). Similar expressions exist for other processes, e.g., 
\begin{eqnarray}
{\cal B} (\mu \to e_L \bar e_L e_L) &=& \frac{v^4}{\Lambda_{\rm LFV}^4} \Bigl[ 2 \left| \vec C \cdot \hat e_{VL} + 4e \vec C \cdot \hat e_{D}\right|^2 + e^2 \left(16 \log \frac{m_\mu^2}{m_e^2} - 68\right) 
\left|\vec C \cdot \hat e_{D}\right|^2\Bigr] 
\nonumber \\
&=& \frac{v^4}{\Lambda_{\rm LFV}^4} \vec C^\dagger {\bf R}_{\mu \to e_L \bar e_L e_L} \vec C < {\cal B}^{\rm exp}_{\mu \to eee}.
\end{eqnarray}
(see \cite{Davidson:2022nnl} for other expressions), with matrices ${\bf R}$ being different for each process. It is then possible to see that other processes are complementary to $\mu \to e \gamma$ and to each other. As described by the Lagrangian (\ref{EffLagFCNC}), four-fermion interactions with leptons only contribute to $\mu \to eee$, LFV interactions with quarks only contribute to $\mu A \to e A$, while the dipole operator can contribute to all processes. Note, however, that the situation changes at one loop. The complementarity of two different processes $A$ and $B$ can be interpreted geometrically as the misalignment between the corresponding vectors in the coefficient space, defined as some angle $\eta$, which can be written in terms of the matrices ${\bf R}$ as 
\beq
\cos^2\eta \sim \frac{\mbox{Tr}\left[{\bf R}_A {\bf R}_B\right]}{\mbox{Tr}\left[{\bf R}_A\right] \mbox{Tr}\left[{\bf R}_B\right]}
\eeq
which vanishes for perfectly complementary processes \cite{Davidson:2022nnl}.

In addition, $A_L$ and $A_R$ can be determined by matching the decay amplitude of Eq.~(\ref{MuEGamma}) to the one derived directly from the SMEFT Lagrangian. The corresponding operators are defined similarly to Eq.~(\ref{MM_SMEFT}), but with different lepton fields, as indicated by the label of the Wilson coefficients,
\beq
A_R = A_L^* = \sqrt{2} \ \frac{v m_i^2}{\Lambda^2} \left(c_W C_{e B}^{fi} - s_W C_{e W}^{fi}\right)  
\equiv \sqrt{2} \ \frac{v m_i^2}{\Lambda^2}  C_\gamma^{fi}
\eeq
where $s_W$ and $c_W$ are the sine and cosine of the Weinberg angle. A bound on ${\cal B}  (\mu \to e\gamma)$ directly constrains a combination of SMEFT Wilson coefficients $C_{e B}^{e\mu}$ and $C_{e W}^{e\mu}$. Based on the latest experimental data, the NP scales probed in such transitions range from $10^5 - 10^7 \GeVcc$, much higher than the scales directly tested at the LHC.

\subsection{Muon decays}
\label{Sec:CLFVDecays}
The $\mu^+ \rightarrow e^+ \gamma$ decay was first investigated by Hincks \& Pontecorvo in 1948~\cite{Hincks:1948vr}, stopping muons in a graphite target and searching for back-to-back emission of a positron and a photon with energies of $\sim52.8 \MeV$ in the muon rest frame. Modern experiments follow the same approach: stopping a positive muon beam in a thin target and seeking the distinctive decay signature. Positive muons are usually chosen to probe decays as they do not undergo nuclear capture, which often leads to the emission of additional particles, introducing extra noise in the detector (and further contamination in the beam when the muons are produced from pion decays). The $\pi^+$ production rate is also larger in proton collisions, increasing the $\mu^+$ yield. Moreover, intense positive muon beams can be produced by the decay of stopped $\pi^+$ near the surface of a production target. The resulting surface muon beam is almost mono-energetic, with a mean energy of $\sim29\MeV$, and can be stopped in thin targets.

Searches for $\mu^+ \rightarrow e^+ \gamma$ must contend with two main sources of background: (i) an intrinsic component from radiative muon decays $\mu^+ \rightarrow e^+ \gamma \nu_e \bar\nu_\mu$, where the neutrinos carry a small amount of energy, and (ii) the accidental coincidence of a positron produced by a regular Michel decay with an energetic photon coming from a radiative muon decay, bremsstrahlung, or positron annihilation in the detector material. The intrinsic contribution scales linearly with the muon stopping rate $R_\mu$, whereas the accidental component has a quadratic dependence, reflecting the combined probabilities of two muon decays. This quadratic dependence has important implications for experimental design. First, the accidental background is minimized with a continuous muon beam instead of a pulsed one, since it has a lower instantaneous rate for a given number of stopped muons. Second, the accidental component becomes dominant as the stopping rate increases. This implies an optimal rate beyond which the signal sensitivity ($S/\sqrt{B}$) no longer increases. Further gain can only be achieved by improving the detector performance, particularly the photon energy and position resolutions, since they contribute quadratically to the background level. Two strategies have been investigated: perform a calorimetric measurement or convert the photon and track the outgoing $e^+e^-$ pair. Photon conversions provide superior energy and angular resolution, but the rate is much reduced compared to the calorimetric option. The converter material also degrades the positron measurement.

\begin{figure}[ht]
    \includegraphics[width=\textwidth]{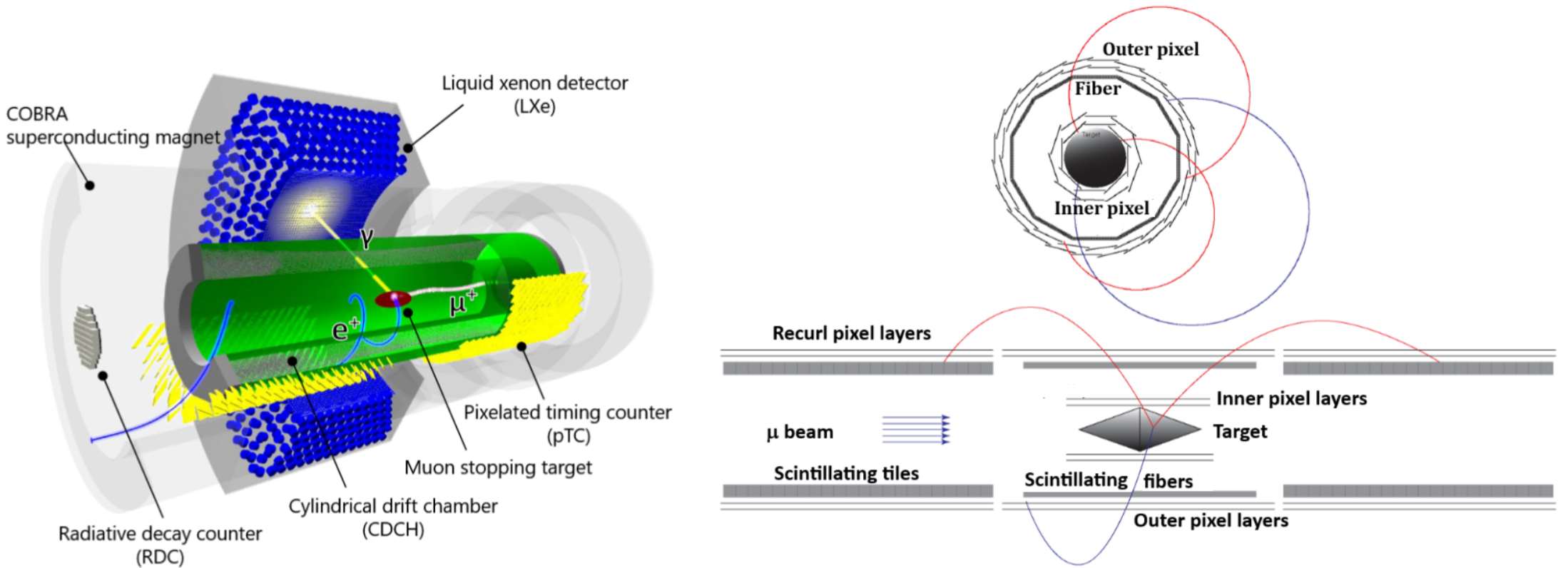}
    \caption{Left: schematic view of the MEG-II experiment together with a simulated event. Right: the Mu3e experiment in the phase I configuration. Figures adapted from References~\cite{MEGII:2023ltw,Mu3e:2020gyw} (CC-BY-NC-ND).}
    \label{fig:megIImu3e}
\end{figure}

The most sensitive $\mu^+ \rightarrow e^+ \gamma$ search to date has been conducted by the MEG II experiment at PSI using the calorimetric approach~\cite{MEGII:2018kmf}. The detector is illustrated schematically in Figure~\ref{fig:megIImu3e}. At its heart is a low-mass cylindrical drift chamber optimized to reconstruct low-energy positrons. The drift chamber is surrounded by a superconducting solenoid generating a graded magnetic field along the axial direction, set so that signal positrons follow a constant projected bending radius almost independently of the emission angle. The photon is measured by a C-shaped liquid xenon calorimeter. A radiative decay counter is placed downstream of the drift chamber to identify low-energy positrons emitted in radiative muon decays, thereby vetoing the energetic photon produced in coincidence to reduce accidental backgrounds. The best bound has been derived with a sample of $3.5 \times 10^{14}$ stopped muons, setting a limit on the branching ratio $\rm BR(\mu^+ \to e^+ \gamma) < 1.5 \times 10^{-13}$ (90\% CL)~\cite{MEGII:2023ltw}. The experiment will continue taking data until 2026 to reach its target sensitivity of $\sim 6 \times 10^{-14}$.

A novel detector concept has been proposed to fully exploit muon beam intensities of $10^9 - 10^{10} \mu^+$/s expected at next-generation beamlines~\cite{Cattaneo:2025bnk}. The design is based on a conversion technique using layers of dense material to convert photons into $e^+e^-$ pairs, subsequently tracked in a magnetic spectrometer. To reduce uncertainties due to energy loss fluctuations in passive material, active converters made of thin layers of LYSO crystals read out by silicon photomultipliers are envisioned. Preliminary studies indicate that $\mu^+ \rightarrow e^+ \gamma$ branching fractions down to $\sim 10^{-15}$ could be achieved. 

The $\mu^+ \rightarrow e^+ e^+e^-$ decay is identified by the presence of two positrons and one electron originating from the same position with a total invariant mass equal to the muon rest mass. Unlike $\mu^+ \rightarrow e^+ \gamma$ decays, the energy of each outgoing particle is not fixed, but continuously distributed up to $m_\mu/2$. The background falls into two categories: internal conversions and accidental combinations. The former arises from $\mu^+ \rightarrow e^+ e^+e^- \nu_\mu \bar\nu_e$ decays (with a branching fraction of $3.4\times 10^{-5}$~\cite{SINDRUM:1985vbg}) where the two neutrinos carry little energy. The total energy of the three particles is similar to the Michel spectrum, but extends up to the muon mass with a long tail~\cite{Djilkibaev:2008jy}. The accidental background is due to the overlay of one positron from standard Michel decays with an $e^+e^-$ pair or two Michel positrons with an electron. Electron-positron pairs arise from photon conversion or Bhabha scattering in the detector, whereas single electrons can be produced by Compton scattering of photons or misreconstructed tracks. As in the $\mu^+ \rightarrow e^+ \gamma$ case, the sensitivity is ultimately limited by the detector performance.

The best limit $\rm BR(\mu^+ \rightarrow e^+ e^+e^-) < 1.0 \times 10^{-12}$ 90\% (CL) has been set by the SINDRUM experiment at PSI~\cite{SINDRUM:1987nra}. The Mu3e proposal at PSI~\cite{Mu3e:2020gyw} aims to extend the sensitivity by four orders of magnitude. The detector is schematically depicted in Figure~\ref{fig:megIImu3e}. The surface muon beam is stopped in a thin double-cone target located in the center of the apparatus. The electron momenta are determined by a cylindrical tracker comprising four (two) layers of silicon pixel sensors in the central (outer) regions, located inside a solenoid magnet. Custom High Voltage Monolithic Active Pixel Sensors (HV-MAPS) thinned down to 50 $\mu$m~\cite{Peric:2007zz} are used to minimize multiple scattering. Including readout components and support structures, each layer has a thickness of only 0.12\% of a radiation length. Optimal momentum resolution is achieved by letting the particles re-curl in the strong magnetic field, crossing the central or outer tracker stations multiple times. Two detectors provide precise timing measurements: a scintillating fiber detector between the inner and outer layers of the central tracker and a thicker scintillating tile device inside the outer layers, where multiple scattering is not a concern. A first physics run with the central part of the detector is foreseen in 2026 to probe branching fractions at the level of $2 \times 10^{-15}$. A second phase will take advantage of the High-Intensity Muon Beam upgrades at PSI~\cite{Aiba:2021bxe} after 2028 to reach the target sensitivity. 

Beyond the canonical searches, muon decay experiments offer sensitivity to processes of the form $\mu^+ \rightarrow e^+ X$ or $\mu^+ \rightarrow e^+ \gamma X$, where $X$ denotes a new light boson that either escapes detection or decays into photons or electrons. In the case of invisible $X$ decays, the $\mu^+ \rightarrow e^+ X$ transition yields a single mono-energetic positron with an energy defined by the $X$ mass, $m_X$. For sufficiently massive $X$, the signal manifests as a distinct peak above the Michel spectrum, detectable with standard bump-hunting techniques. The situation is more challenging for (nearly) massless $X$, as the positron line becomes essentially indistinguishable from the endpoint of the Michel spectrum. Fortunately, current surface muon beams are nearly 100\% polarized, a feature that can be leveraged to suppress background. In the SM, polarized muon decay vanishes at$\cos(\theta_e) = 1$ (where $\theta_e$ is the angle between muon and positron momenta). For $0 < \cos(\theta_e) <1$, polarization shifts the Michel spectrum away from the mono-energetic signal. Consequently, restricting measurements to $\cos(\theta_e) \sim 1$ significantly enhances sensitivity for low-mass X searches. This approach has been proposed in the MEG-II-fwd concept, an extension of the MEG-II experiment to search for axion-like particles (ALP)~\cite{Calibbi:2020jvd}. 

\begin{figure}[ht]
    \centering   
    \includegraphics[width=0.85\textwidth]{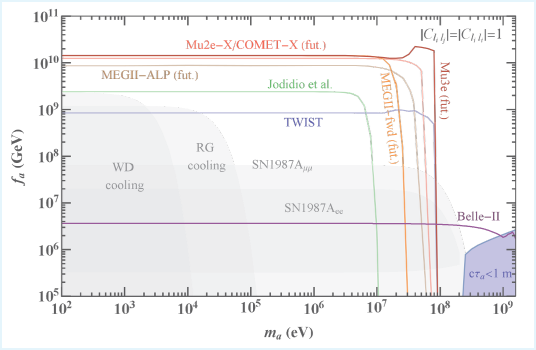}
    \caption{Current and future limits on the axion-like particle decay constant, $f_a$, with identical couplings to all generations of leptons. The gray shaded regions are excluded by the astrophysical bounds from star cooling, while the parameter space corresponding to displaced decays potentially accessible to detectors ($c\tau_a < 1 \rm \, m$) is depicted in blue. The green solid line represents the bounds derived from a search for $\mu^+ \rightarrow e^+ a$ by Jodidio {\it et al.}~\cite{Jodidio:1986mz}. The expected reach of future muon decays and conversion experiments is shown as dashed lines. For more details, see References~\cite{Calibbi:2020jvd, Hill:2023dym, Knapen:2023zgi}. Figure adapted from Reference~\cite{Hill:2023dym} (CC BY 4.0), which was adapted from Reference~\cite{Calibbi:2020jvd} (CC BY 4.0).}
    \label{fig:ALPexp}
\end{figure}

Figure~\ref{fig:ALPexp} illustrates the projected reach of the MEGII-fwd setup, alongside existing constraints from previous muon decay experiments and astrophysical observations~\cite{Calibbi:2020jvd,Hill:2023dym}. Meanwhile, the Mu3e experiment intends to leverage real-time track reconstruction via its filter farm to search for signatures of New Physics. An ALP would appear as a peaking excess above the smooth background in the track momentum spectrum. These experiments could probe ALP couplings $f_a \sim 10^{10} \GeVcc$, two orders of magnitude over present bounds for ALP masses $m_{a} \lesssim 90 \MeVcc$. Perspectives on complementary ALPs scenarios and new dark gauge bosons decaying invisibly are further discussed in Reference~\cite{Hill:2023dym}.

Visible final states can be investigated in $\mu^+ \rightarrow e^+ X$ followed by $X \rightarrow \gamma \gamma$ or $X \rightarrow e^+e^-$ for both prompt and displaced $X$ decays. The process $\mu^+ \rightarrow e^+ \gamma \gamma$ has been measured by the Crystal Box detector~\cite{Bolton:1988af} and the MEG experiments~\cite{MEG:2020zxk}, probing branching fractions in the range ${\cal O}(10^{-11} - 10^{-10})$. The improved granularity of the MEG II calorimeter, combined with higher statistics, is expected to significantly increase the sensitivity in the coming years. Mu3e is ideally suited to investigate $X \rightarrow e^+e^-$ final states with either prompt or displaced decays. The background can be significantly suppressed if the particle travels at a measurable distance inside the hollow target. For example, lepton flavor violating and conserving ALP couplings as small as $\sim 10^{-15}$ could be probed, complementing searches at other experiments~\cite{Knapen:2024fvh}. Light dark photons, hypothetical gauge bosons coupling the dark sector to the SM photon via kinetic mixing, could also be radiated in muon decays and reconstructed as an $e^+e^-$ resonance~\cite{Echenard:2014lma, Hesketh:2022wgw}. 

\subsection{Muon conversion}

The $\mu^-N - e^-N$ conversion process is studied by stopping negative muons in a target, where they are captured to form muonic atoms and rapidly cascade down to the ground state. Once captured, the muon can either decay in orbit ($\mu^- \rightarrow e^- \bar\nu_e \nu_\mu$), undergo nuclear capture ($\mu^- + (Z,A) \rightarrow \nu_\mu + (Z-1,A)$), or convert to an electron without the emission of additional neutrinos ($\mu^-N \rightarrow e^-N$). Only decays in orbit and nuclear captures are allowed in the Standard Model. Their relative rates depend on the nuclear structure, the capture rate generally increasing with the atomic number Z ({\it e.g.}, the capture rate is $\sim 61\%$ in Al and $\sim 95\%$ in Au~\cite{Suzuki:1987jf}). While the energy spectrum of muon decay in orbit resembles that of free muon decays below the Michel edge, the outgoing electron can exchange momentum with the nucleus, generating a tail extending up to the energy of the conversion process~\cite{Szafron_2016, Fontes:2025mps}.

Unlike the SM processes, muon-to-electron conversion is a coherent process, \ie, the nucleus remains unchanged, and the outgoing electron has a well-defined energy 
$$E_e = m_\mu -E_{b,1S} - E_{recoil}$$ 
where $E_{b,1S}$ and $E_{recoil}$ denote the muon binding energy and the recoil energy of the nucleus, respectively. For $^{27}$Al, the electron energy is $104.9 \MeV$. The rate $R_{\mu e}$ is conventionally defined as the ratio between the $\mu - e$ conversion rate and the total muon capture rate:
$$R_{\mu e} = \frac{\Gamma(\mu^- + N(Z,A) \rightarrow e^- + N(Z,A))} {\Gamma(\mu^- + N(Z,A) \rightarrow \rm all \, captures)}$$ 
where $N(Z,A)$ denotes the atomic number and mass of the target nuclei. In contrast to decay experiments, the experimental sensitivity is only limited by the ability to reject the intrinsic and cosmic-induced backgrounds; no accidental contribution is expected.

\begin{figure}[ht]
    \centering   
    \includegraphics[width=\textwidth]{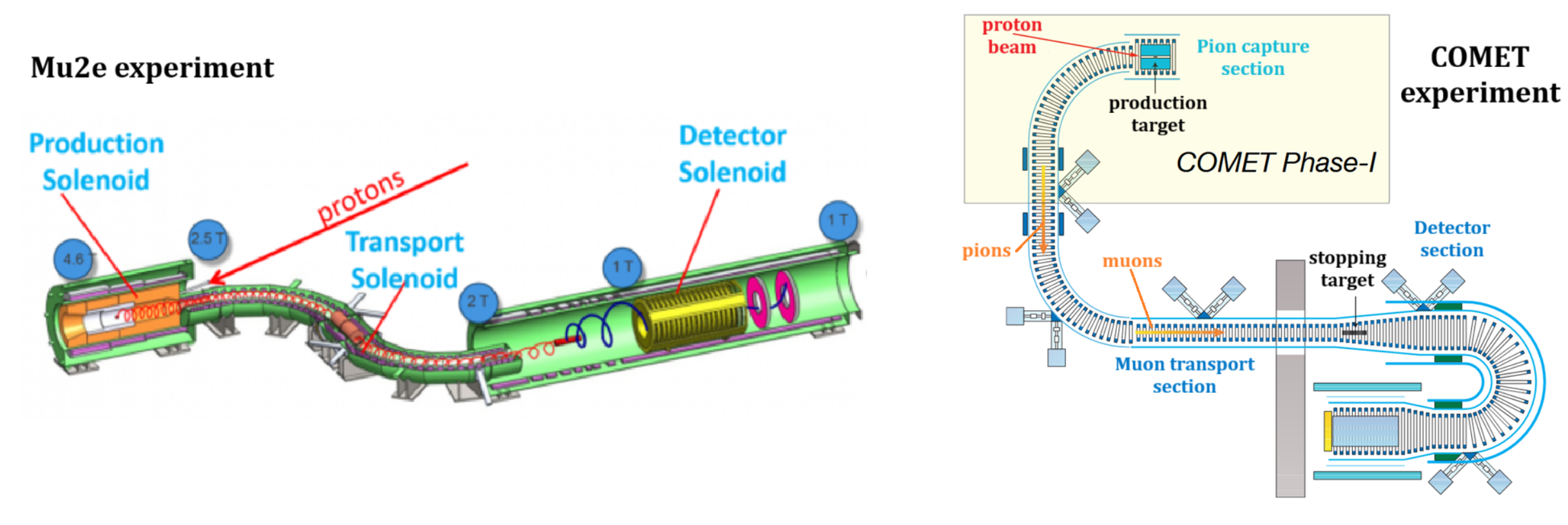}
    \caption{Left: schematic views of the Mu2e experiment at Fermilab. Right: illustration of the COMET experiment at J-PARC. Figures adapted from References~\cite{bartoszek2015mu2e, COMET:2009qeh} (arXiv nonexclusive-distrib. 1.0).}
    \label{fig:mu2ecomet}
\end{figure}

The current experimental bound on muon-to-electron conversion was established by the SINDRUM-II experiment at PSI using an Au target, $R_{\mu e} < 7\times10^{-13}$ at 90\% CL~\cite{SINDRUMII:2006dvw}. The Mu2e experiment at Fermilab~\cite{bartoszek2015mu2e} and the COMET experiment at J-PARC~\cite{COMET:2009qeh, Krikler:2016qij}, both under construction, aim to improve upon this sensitivity by four orders of magnitude. Both experiments exploit a scheme originally proposed for the MELC experiment~\cite{osti_5665919} to overcome the limitations due to beam backgrounds. Since pions have a much shorter lifetime than low-$Z$ muonic atoms and decay shortly after the arrival of the proton pulse on the target, a pulsed beam allows for delaying the search for the conversion signal until pion-related backgrounds are effectively suppressed. The Mu2e and COMET experimental setups, shown in Figure~\ref{fig:mu2ecomet}, comprise three main sections. The muon beam is created in the production section from the decay of pions produced in the collision of pulsed proton bunches in a production target. The target is surrounded by a strong magnetic field, increasing the muon capture probability by several orders of magnitude compared to conventional muon facilities~\cite{Hino:2014bpx}. The graded field directs slow backward muons into the curved transport section. Positive and negative particles drift in opposite vertical directions when propagating in a curved solenoidal magnetic field, allowing the selection of negative (or positive) muons with a collimator (for Mu2e) or by applying a corrector field to push negative muons back onto the central magnet axis (for COMET). Finally, the detector section comprises an aluminum stopping target to capture muons and a detector system to identify the conversion electron. 

In the Mu2e configuration, the stopping target is positioned in front of the detectors. The momentum of charged particles is determined primarily through a low-mass proportional straw tube
tracker with a resolution of $\sim 180 \keVc$ at $105\MeVc$. A calorimeter is located downstream of the tracker to provide complementary information for particle identification, track reconstruction, and fast triggering. While this design is charge symmetric, the innermost region of the detector must be left uninstrumented to withstand the considerable flux of low-transverse momentum particles from muon decays and beam remnants. COMET will be carried out in two stages. In the first phase, a stopping target is placed in the first portion of the transport solenoid, surrounded by a cylindrical drift chamber and two arrays of trigger hodoscope. An additional curved solenoid downstream of the stopping target is installed in the second phase to reject neutral and low-momentum charged particles. The drift chamber is replaced by a low-mass proportional straw tube tracker and a calorimeter. This scheme allows for a fully instrumented volume, but at the cost of charge symmetry. In both experiments, an active veto system rejects cosmic ray backgrounds, and dedicated stopping target monitors measure the characteristic X-ray emitted by muon cascades or nuclear capture to estimate the number of stopped muons. 

Mu2e plans to collect data during two running periods separated by a pause to upgrade the PIP-II complex. The single event sensitivity for the first run is $R_{\mu e} = 2.4 \times 10^{-16}$ on Al~\cite{Mu2e:2022ggl}, with an order of magnitude improvement anticipated at the end of the second data taking period. COMET is expected to reach similar performance levels for Phase-I (${\cal O}(10^{-15})$) and Phase-II (${\cal O}(10^{-17})$). Assuming ${\cal O}(1)$ Wilson coefficients, these bounds would probe NP mass scales $\Lambda \sim 10^3 - 10^4 \TeVcc$.

In addition to the primary conversion search, Mu2e and COMET Phase-I can study the $\mu^- + N(Z,A) \rightarrow e^+ + N'(Z-2,A)$ conversion, a reaction violating both lepton number ($\Delta L=2$) and lepton flavor. This process is complementary to neutrinoless double beta decays ($0\nu \beta\beta$), with enhanced sensitivity to BSM models with enhanced "flavor off-diagonal" transitions. The Black Box Theorem~\cite{Schechter:1981bd} also implies that a non-zero Majorana neutrino mass or New Physics must be at play if this process is observed~\cite{Hirsch:2006yk}. The most stringent bound on $\mu^- - e^+$ conversion has been set by SINDRUM-II, BR($\mu^- + {\rm Ti} \rightarrow e^+ + {\rm Ca}) < 1.7\times 10^{-12}$ at 90\% CL~\cite{SINDRUMII:1998mwd}. Projections for Mu2e and COMET Phase-I suffer from large uncertainties on the knowledge of radiative muon captures, the main background, but preliminary results indicate that an improvement of up to four orders of magnitude could be realized~\cite{Lee:2021hnx}. 

Mu2e and COMET could also collect enormous samples of positive muon and pion decays, enabling searches for light new particles with lepton flavor violating couplings~\cite{Hill:2023dym}. The case of axion-like particles has been reviewed in Section~\ref{Sec:CLFVDecays}; the reach of Mu2e and COMET is comparable to that of muon decay experiments. Heavy neutral leptons ($N$) could also be studied in $\pi^+ \rightarrow e^+ N$ decays with unprecedented sensitivity in the low-mass region~\cite{Huang:2022xii}, complementing the reach of the dedicated PIONEER experiment~\cite{PIONEER:2022yag}. 
 
On a more distant horizon, Mu2e-II~\cite{Mu2e-II:2022blh} has been proposed as an evolution of the Mu2e experiment, aiming to increase the sensitivity to $\mu^- - e^-$ conversion by an order of magnitude beyond Mu2e by leveraging the PIP-II linear accelerator at Fermilab~\cite{Ball2017}. If a conversion signal is detected, Mu2e-II could explore the underlying physics by measuring conversion rates using mid-$Z$ target materials like titanium or vanadium~\cite{Cirigliano:2009bz, Borrel:2024ylg}. Should no signal emerge from current experiments, pushing the sensitivity further to explore higher NP mass scales will be essential. 

Beyond that, the Advanced Muon Facility (AMF) has been envisioned as a next-generation facility to perform muon physics with unparalleled sensitivity~\cite{CGroup:2022tli} by fully leveraging the capabilities of the PIP-II accelerator. Inspired by the PRISM architecture~\cite{KUNO2005376}, AMF utilizes a compact fixed-field alternating (FFA) gradient synchrotron to produce a pure, monochromatic muon beam.  
This facility would support a diverse experimental program, including charged lepton flavor violation, muonium-antimuonium oscillation studies, searches for the muon electric dipole moment, and muon spin rotation measurements. For example, CLFV experiments at AMF could potentially enhance sensitivity by up to two orders of magnitude beyond current projections, achieving conversion rates as low as $10^{-19}$ with multi-MW beams and probing NP mass scales up to $10^4 - 10^5 \TeVcc$.

\subsection{Muonium - antimuonium transitions}
Muonium is a pure QED bound state of a positively charged muon and a negatively charged electron, $\ket{M} \equiv \ket{\mu^+e^-}$. A $\Delta L_\mu=2$ interaction could convert the muonium state into the antimuonium state, opening the possibility of muonium-antimuonium oscillations \cite{Pontecorvo:1957cp, Feinberg:1961zza}. As a variety of well-motivated NP models include $\Delta L_\mu=2$ interaction terms\footnote{Strictly speaking, in this section, we are interested in processes with $\Delta L_\mu=-\Delta L_e=2$, with $\Delta L=\Delta L_\mu+\Delta L_e=0$. Here, we will refer to these processes simply as $\Delta L_\mu=2$. See also \cite{Heeck:2024uiz}.}, observing muonium to antimuonium conversion would be a clear signal of NP in the leptonic sector. A model-independent calculation of the oscillation parameters using effective theory techniques is feasible, expressing matrix elements as a series of local operators expanded in inverse powers of $m_\mu$ \cite{Conlin:2020veq}.
 
If the NP Lagrangian includes $\Delta L_\mu=2$ lepton flavor violating interactions, the time evolution of muonium and antimuonium states becomes coupled, and it is appropriate to consider their combined evolution, which is governed by a Schr\"odinger-like equation  
\begin{equation}\label{two state time evolution}
i\frac{d}{dt}
\begin{pmatrix}
\ket{M(t)} \\ \ket{\overline{M}(t)}
\end{pmatrix}
=
\left(m-i\frac{\Gamma}{2}\right)
\begin{pmatrix}
\ket{M(t)} \\ \ket{\overline{M}(t)}
\end{pmatrix}
\end{equation}
where $m$ and $\Gamma$ are $2\times2$ matrices. $\CPT$ invariance requires that the masses and widths of muonium and antimuonium are equal, so $m_{11} = m_{22}$ and $\Gamma_{11}=\Gamma_{22}$. Assuming $\CP$ invariance of the $\Delta L_\mu = 2$ interaction for simplicity further dictates that $m_{12}=m^{*}_{21}$ and $\Gamma_{12}=\Gamma^{*}_{21}$. The Hamiltonian in Eq.~(\ref{two state time evolution}) is not Hermitian since the states can decay. The mass eigenstates $ |\mmu_{1,2} \rangle$ can be defined as
\beq
 |\mmu_{1,2} \rangle = \frac{1}{\sqrt{2}} \left[ |\mmu \rangle \mp  |\ammu \rangle
 \right] 
\eeq
where $\CP$ violation is neglected and the convention $\CP |\mmu_\pm 
\rangle = \mp  |\mmu_\pm \rangle$ is used. The mass and width differences of the mass eigenstates are 
\begin{eqnarray}\label{XandY}
x = \frac{\Delta m}{\Gamma} = \frac{M_{1}-M_{2}}{\Gamma}, \qquad y =  \frac{\Delta \Gamma}{2\Gamma} \equiv \frac{\Gamma_{2}-\Gamma_{1}}{2\Gamma}
\end{eqnarray} 
where  $M_i$ ($\Gamma_i$) are the masses (widths) of the mass eigenstates $ |\mmu_{1,2} \rangle$,  the average lifetime $\Gamma=(\Gamma_1+\Gamma_2)/2$, and $\Delta m$ and $\Delta \Gamma$ can be either positive or negative. While $\Gamma$ is dominated by the SM decay rate of the muon, $x$ and $y$ require lepton flavor violating interactions, and it is thus expected that both $x,y \ll 1$. 

Denoting the amplitude for muonium decay into a final state $f$ as $A_f = \langle f|{\cal H} |\mmu\rangle$ and the amplitude for its decay into a $\CP$-conjugate final state $\overline{f}$ as $A_{\bar f} = \langle \overline{f}|{\cal H} |\mmu\rangle$, the time-dependent decay rate of $\mmu$ into $\overline{f}$ can be written as
\beq
\Gamma(\mmu \to \overline{f})(t) = \frac{1}{2} N_f \left|A_f\right|^2 e^{-\Gamma t} \left(\Gamma t\right)^2 R_M(x,y), \ \ \mbox{with} \ \ R_M(x,y) = \frac{1}{2} \left(x^2+y^2\right),
\eeq
where $N_f$ described a phase-space factor and $R_M(x,y)$ is the oscillation rate \cite{Petrov:2021idw}. Integrating over time and normalizing to $\Gamma(\mmu \to f)$, the probability of $\mmu$ decaying as $\ammu$ at time $t > 0$ is given by
\beq\label{Prob_osc}
P_{M \bar{M}} = P(\mmu \rightarrow \ammu) = \frac{\Gamma(\mmu \to \overline{f})}{\Gamma(\mmu \to f)} = R_M(x,y). 
\eeq
Note that Eq.~(\ref{Prob_osc}) depends on both the mass $x$ and the lifetime $y$ differences \cite{Conlin:2020veq}. 

The Standard Model contribution to Eq.~(\ref{Prob_osc}) is tiny, making muonium-antimuonium oscillation a valuable probe of New Physics \cite{Petrov:2021idw}. If NP particles are heavy, an effective Lagrangian, represented by a collection of effective operators changing the lepton quantum number by two units, can be constructed
\begin{equation}\label{DL2}
{\cal L}_{\rm eff}^{\Delta L_\mu=2}=-\frac{1}{\Lambda^{2}}\underset{i}{\sum}C^{\Delta L=2}_{i}(\mu)Q_{i}(\mu)
\end{equation}
with the operators built entirely from the muon and electron degrees of freedom,
\begin{eqnarray}\label{Dim6_Op}
Q_1 &=& \left(\overline\mu_L \gamma_\alpha e_L \right) \left(\overline\mu_L \gamma^\alpha e_L \right), \quad
Q_2 = \left(\overline\mu_R \gamma_\alpha e_R \right) \left(\overline\mu_R \gamma^\alpha e_R \right),
\nonumber \\
Q_3 &=& \left(\overline\mu_L \gamma_\alpha e_L \right) \left(\overline\mu_R \gamma^\alpha e_R \right), \quad	
Q_4 = \left(\overline\mu_L e_R \right) \left(\overline\mu_L e_R \right), \quad
Q_5 = \left(\overline\mu_R e_L \right) \left(\overline\mu_R e_L \right).
\end{eqnarray}
A $\Delta L_\mu=2$ interaction described by the Lagrangian~(\ref{DL2}) can both lead to muon decays, such as $\mu^+ \to 3e$, and to muonium--antimuonium oscillations. In addition to these operators, other $\Delta L_\mu=2$ operators containing muon, electron, and neutrino fields can be constructed \cite{Conlin:2020veq},
\beq\label{Dim6_Op_nu}
Q_6 = \left(\overline\mu_L \gamma_\alpha e_L \right) \left(\overline{\nu_{\mu}}_L \gamma^\alpha {\nu_e}_L \right), \quad
Q_7 = \left(\overline\mu_R \gamma_\alpha e_R \right) \left(\overline{\nu_{\mu}}_L \gamma^\alpha {\nu_e}_L \right),
\eeq
where operators containing only left-handed neutrinos are included. The operators in Eq.~(\ref{Dim6_Op_nu}) could lead to both muon decays $\mu \to e \nu_\mu \bar \nu_e$ and the lifetime difference in muonium--antimuonium oscillations.

%
%
Assuming that only one operator at a time provides a dominant contribution (the single operator dominance hypothesis), it is possible to constrain the Wilson coefficients of each operator. The constraints on the NP scale can be derived from data obtained by the MACS experiment. These constraints, while different for the various chiral structures of the operators in Eq.~(\ref{Dim6_Op}), are all at the few-$\TeV$ level~\cite{Conlin:2020veq}. This is due to different magnetic-field suppression factors. In contrast, constraints on the NP scales from the operators in Eq.~(\ref{Dim6_Op_nu}) are significantly less stringent, at the $10^{-4} \TeVcc$ level. It is worth mentioning that this assumption is {\it not} realized in many specific ultraviolet completions of the LFV EFTs, so cancellations among contributions from different operators to $x$ and $y$ are possible. Despite this shortcoming, the single operator dominance approach remains a valuable tool for constraining the Lagrangian parameters.

Experimentally, muonium-to-antimuonium conversion is generally investigated by producing muonium from a surface muon injected into a suitable material. The muons are slowed down, and a fraction undergoes spontaneous electron capture, forming muonium that subsequently diffuses into the vacuum. The conversion process is identified by detecting both the electron and positron released by antimuonium decay. An energetic electron is produced by the $\mu^- \rightarrow e^- \bar\nu_e \nu_\mu$ with a characteristic energy spectrum extending up to the Michel edge, together with a positron emitted from the atomic shell with a mean energy of $13.5 \eV$. The MACS experiment at PSI exploited this technique to establish a bound on the $M - \bar{M}$ conversion probability ($P_{M \bar{M}}$) of $P_{M \bar{M}} < 8.3 \times 10^{-11}$ at 90\% CL~\cite{Willmann:1998gd}. 

The proposed MACE experiment~\cite{Bai:2024skk} aims to further enhance the sensitivity by over two orders of magnitude using the China initiative Accelerator Driven System (CiADS)~\cite{Cai:2023caf} to produce an intense surface muon beam. A major improvement resides in the development of a perforated silica aerogel target to enhance the production and diffusion of muonium atoms into the vacuum~\cite{Beare:2020gzr}. The detector layout is conceptually similar to MACS. The Michel electron is identified by a magnetic spectrometer surrounding the target. It consists of a cylindrical drift chamber and a set of tiled timing counters immersed in a magnetic field. A positron transport system guides low-energy positrons to a micro-channel plate detector located downstream while conserving their transverse position. The back-to-back photons produced by the positron annihilation inside the detector material are reconstructed by a high-resolution crystal calorimeter. The main backgrounds arise from accidental coincidences of energetic electrons produced by Bhabha scattering of positrons emitted by muonium decays, and $\mu^+ \rightarrow e^+e^-e^+ \nu_e \bar\nu_\mu$ decays in which an energetic electron and a slow positron are produced while the other positron remains undetected. Coincidence between the Michel electron, the positron signal, and the calorimeter-detected photon pair is expected to reduce the background level below one count for $10^{14}$ muonium decay in vacuum, probing a transition probability $ P_{M \bar{M}} \sim 10^{-13}$. Such sensitivity could explore $\Delta L=2$ processes at the scale of $10-100 \TeVcc$~\cite{Bai:2024skk}.

\section{LOOKING FORWARD}
Almost a century after its discovery, the muon remains one of the most sensitive and versatile probes of new physics. Thanks to its relatively long lifetime and the availability of intense sources, the muon offers a wide range of experimental avenues to explore new phenomena and, more broadly, the questions of flavor and generations. This review has highlighted recent advances in precision muon physics over the last decade, summarized in Table~\ref{summaryExp}. 

The recent measurement of the muon anomalous magnetic moment has confirmed with astonishing precision the SM predictions based on perturbative QED and lattice QCD calculations, setting strong constraints on the presence of light new degrees of freedom. Additional efforts are underway to understand the discrepancy with the theoretical value estimated from data-driven measurements, and new experiments are being developed to provide an independent measurement of the magnetic anomaly. The study of charged lepton flavor violating processes promises impressive gains in the next decade, both for muon decays and conversion to electrons. Current bounds already probe mass scales well above that directly accessible by colliders, and planned experiments will further increase the sensitivity by an order of magnitude. Models of light new physics, including lepton flavor violating axions and hidden sector particles, could also be explored with unprecedented accuracy. 

Future muon experiments and new dedicated facilities will enable another leap in sensitivity in the coming decades. Should a signal be observed, a comprehensive series of measurements in the muon sector would provide valuable insights into the nature of the underlying physics and the question of flavor. We can only hope that these perspectives will convince the reader of the importance and bright future of precision muon physics. 

\begin{table}
\centering
\small
\begin{tabular}{|l|l|l|l|}
\hline 
Experiment     & Status  & Physics sensitivity & Ref. \\\hline
TWIST                        & Completed    &$\Delta\rho, \Delta \delta, \Delta(P^\pi_\mu \xi) \sim 10^{-4}$ & \cite{TWIST:2011aa, TWIST:2011egd}\\
Muon g-2 (Fermilab)              & Completed    & $\Delta a_\mu / a_\mu$ - 127 ppb & \cite{Muong-2:2025xyk}\\
Muon g-2/EDM (J-PARC)        & Prototyping  & $\Delta a_\mu / a_\mu$ - 450 ppb & \cite{Abe:2019thb}\\
                             &              & $d_\mu \sim 10^{-21} \rm \, e \cdot cm$ & \\
muEDM (PSI)                  & Prototyping  & $d_\mu \sim 10^{-23} \rm \, e \cdot cm$ & \cite{Sakurai:2022tbk}\\
MUonE (CERN)                 & Data taking  & $\Delta a_\mu^{\rm HPV} / a_\mu^{\rm HPV} \sim 0.5\%$ & \cite{MUonE:2016hru}\\
MEG-II (PSI)   & Data taking  & ${\cal B}(\mu\rightarrow e\gamma) \sim 6\times 10^{-14}$ & \cite{MEGII:2018kmf}\\
Mu3e (PSI)     & Construction & ${\cal B}(\mu \rightarrow 3e) \sim10^{-15}$ (phase I)    & \cite{Mu3e:2020gyw}\\
COMET (J-PARC) & Construction & $R_{\mu e} \sim 10^{-15}$ (phase I)                      & \cite{COMET:2009qeh}\\
Mu2e (Fermilab)    & Construction & $R_{\mu e} \sim 2.4 \times10^{-16}$ (phase I)                         & \cite{bartoszek2015mu2e,Mu2e:2022ggl}\\
MACE (CSNS)    & R\&D &       $ P(M \rightarrow\bar M) \sim 10^{-13}$                    & \cite{Bai:2024skk}\\
Mu2e-II (Fermilab) & R\&D         & $R_{\mu e} \sim 10^{-18}$                                & \cite{Mu2e-II:2022blh}\\
AMF (Fermilab)     & R\&D         & $R_{\mu e} \sim 10^{-19}$                                & \cite{CGroup:2022tli} \\\hline
\end{tabular}

\caption{An overview of the experiments discussed in this article. Experiments in the R\&D stage are currently undergoing feasibility studies. }
\label{summaryExp}
\end{table}

\section*{DISCLOSURE STATEMENT}
The authors are not aware of any affiliations, memberships, funding, or financial holdings that might be perceived as affecting the objectivity of this review.

\section*{ACKNOWLEDGMENTS}
We thank David Hitlin for useful comments and suggestions to improve the quality of the manuscript. This work was supported by the U.S. Department of Energy, Office of Science, Office of High Energy Physics, under Award Number DE-SC0024357 (AAP) and DE-SC0011925 (BE).



\begin{thebibliography}{96}
\expandafter\ifx\csname
natexlab\endcsname\relax\def\natexlab#1{#1}\fi

\bibitem{Anderson:1936zz} C.~D.~Anderson and S.~H.~Neddermeyer, Phys. Rev. \textbf{50}, 263 (1936)
\bibitem{Yukawa:1935xg} H.~Yukawa, Proc. Phys. Math. Soc. Jap.  \textbf{17}, 48 (1935)
\bibitem{Conversi:1945qhg} M.~Conversi, E.~Pancini and O.~Piccioni, Phys. Rev  \textbf{68}, 232 (1945)
\bibitem{Conversi:1947ig} M.~Conversi, E.~Pancini and O.~Piccioni, Phys. Rev  \textbf{71}, 209 (1947)
\bibitem{Hincks:1948vr} E.~P.~Hincks and B.~Pontecorvo, Phys. Rev. \textbf{73}, 257-258 (1948)
\bibitem{Danby:1962nd} G.~Danby, \textit{et al.}, Phys. Rev. Lett  \textbf{9}, 36 (1962)
\bibitem{PhysRev.105.1415} R.~Garwin, L.~Lederman and M. Weinrich, Phys. Rev. \textbf{105}, 1415 (1957)
\bibitem{Gorringe:2015cma} T.~P.~Gorringe and D.~W.~Hertzog, Prog. Part. Nucl. Phys.  \textbf{84}, 73 (2015) 
\bibitem{Keshavarzi:2022kpc} A.~Keshavarzi, K.~S.~Khaw and T.~Yoshioka, Nucl. Phys. B \textbf{975}, 115675 (2022)
\bibitem{Calibbi:2017uvl} L.~Calibbi and G.~Signorelli, Riv. Nuovo Cim.  \textbf{41}, 71 (2018)
\bibitem{Bernstein:2013hba}R.~H.~Bernstein and P.~S.~Cooper, Phys. Rept. \textbf{532}, 27 (2013)
\bibitem{Kinoshita:1958ru} T.~Kinoshita and A.~Sirlin, Phys. Rev. \textbf{113}, 1652-1660 (1959)
\bibitem{Marciano:1988vm} W.~J.~Marciano and A.~Sirlin, Phys. Rev. Lett. \textbf{61}, 1815-1818 (1988)
\bibitem{Petrov:2021idw} A.~A.~Petrov, ``Indirect Searches for New Physics'',  CRC Press, 2021, ISBN 978-1-351-17601-9
\bibitem{Kuno:1999jp} Y.~Kuno and Y.~Okada, Rev. Mod. Phys. \textbf{73}, 151-202 (2001)
\bibitem{Bouchiat:1957zz} C.~Bouchiat and L.~Michel, Phys. Rev. \textbf{106}, 170-172 (1957)
\bibitem{TWIST:2011aa} A.~Hillairet \textit{et al.} [TWIST], Phys. Rev. D \textbf{85}, 092013 (2012)
\bibitem{TWIST:2011egd} J.~F.~Bueno \textit{et al.},  [TWIST], Phys.  Rev. D  \textbf{84}, 032005 (2011)
\bibitem{ParticleDataGroup:2024cfk} S.~Navas \textit{et al.} [Particle Data Group], Phys. Rev. D  \textbf{110}, 030001 (2024)
\bibitem{Herczeg:1985cx} P.~Herczeg, Phys. Rev. D \textbf{34}, 3449 (1986)
\bibitem{Schwinger:1948iu} J.~S.~Schwinger, Phys. Rev. \textbf{73}, 416-417 (1948)
\bibitem{Aoyama:2012wk} T.~Aoyama, M.~Hayakawa, T.~Kinoshita and M.~Nio, Phys. Rev. Lett. \textbf{109}, 111808 (2012)
\bibitem{Aliberti:2025beg} R.~Aliberti, \textit{et al.}, Phys. Rept. \textbf{1143}, 1-158 (2025)
\bibitem{Peskin:1995ev} M.~E.~Peskin and D.~V.~Schroeder,  Addison-Wesley, 1995, ISBN 978-0-201-50397-5, 978-0-429-50355-9, 978-0-429-49417-8
\bibitem{Fortuna:2024rqp} F.~Fortuna, J.~M.~M{\'a}rquez and P.~Roig, Phys. Rev. D \textbf{111}, no.7, 075012 (2025)
\bibitem{Jarlskog:1985ht} C.~Jarlskog, Phys. Rev. Lett. \textbf{55}, 1039 (1985)
\bibitem{Barbieri:1974nc} R.~Barbieri and E.~Remiddi, Nucl. Phys. B \textbf{90}, 233-266 (1975)
\bibitem{Jegerlehner:2009ry} F.~Jegerlehner and A.~Nyffeler, Phys. Rept. \textbf{477}, 1-110 (2009)
\bibitem{CERN-Mainz-Daresbury:1978ccd}J.~Bailey \textit{et al.},  Nucl. Phys. B \textbf{150}, 1 (1979)
\bibitem{Muong-2:2006rrc} G.~Bennett \textit{et al.}, Phys. Rev. D. \textbf{73}, 072003 (2006)
\bibitem{Muong-2:2015xgu} J.~Grange \textit{et al.}, \textit{Muon (g-2) Technical Design Report}, arXiv:1501.06858 (2015)
\bibitem{Muong-2:2025xyk} D.~Aguillard \textit{et al.}, Phys. Rev. Lett. \textbf{135}, 101802 (2025)
\bibitem{MUonE:2016hru} G.~Abbiendi \textit{et al.}, Eur. Phys. J. C \textbf{77}, 139 (2017)
\bibitem{Abe:2019thb} M.~Abe \textit{et al.}, PTEP \textbf{5}, 053C02 (2019) 
\bibitem{Muong-2:2008ebm} G.~Bennett \textit{et al.}, Phys.  Rev. D \textbf{80}, 052008 (2009)
\bibitem{Sakurai:2022tbk} M.~Sakurai \textit{et al.}, JPS Conf. Proc. \textbf{37}, 020604 (2022)
\bibitem{Davidson:2022nnl} S.~Davidson and B.~Echenard, Eur. Phys. J. C \textbf{82}, 836 (2022)
\bibitem{Bilenky:1977du} S.~M.~Bilenky, S.~T.~Petcov and B.~Pontecorvo, Phys. Lett. B \textbf{67}, 309 (1977)
\bibitem{MEGII:2018kmf} A.~Baldini \textit{et al.}, Eur. Phys. J. C \textbf{78}, 380 (2018)
\bibitem{MEGII:2023ltw} K.~Afanaciev \textit{et al.}, Eur. Phys. J. C \textbf{84}, 216 (2024), [Erratum: Eur. Phys. J. C \textbf{84}, 1042 (2024)]
\bibitem{Cattaneo:2025bnk} P.~W.~Cattaneo, \textit{et al.},``Future perspectives for $\mu^+ \to \mathrm{e}^+ \gamma$ searches,'' arXiv:2504.18831
\bibitem{SINDRUM:1985vbg} W.~H.~Bertl \textit{et al.} [SINDRUM], Nucl. Phys. B \textbf{260}, 1-31 (1985)
\bibitem{Djilkibaev:2008jy} R.~Djilkibaev and R.~Konoplich, Phys.  Rev. D  \textbf{79}, 073004 (2009)
\bibitem{SINDRUM:1987nra} U.~Bellgardt \textit{et al.}, Nucl. Phys. B \textbf{299}, 1 (1988)
\bibitem{Mu3e:2020gyw} K.~Arndt \textit{et al.}, Nucl. Instrum. Meth. A \textbf{1014}, 165679 (2021)
\bibitem{Peric:2007zz} I.~Peric, Nucl. Instrum. Meth. A  \textbf{582}, 876 (2007)
\bibitem{Aiba:2021bxe} M.~Aiba, \textit{et al.}, `Science Case for the new High-Intensity Muon Beams HIMB at PSI,'' arXiv:2111.05788
\bibitem{Calibbi:2020jvd} L.~Calibbi, D.~Redigolo, R.~Ziegler and J.~Zupan, JHEP  \textbf{9}, 173 (2021)
\bibitem{Hill:2023dym} R.~Hill, R.~Plestid and J.~Zupan Phys. Rev. D \textbf{109}, 035025 (2024)
\bibitem{Jodidio:1986mz} A.~Jodidio \textit{et al.}, Phys. Rev. D \textbf{34}, 1967 (1986)
\bibitem{Knapen:2023zgi} S.~Knapen, K.~Langhoff, T.~Opferkuch and D.~Redigolo, JHEP \textbf{7}, 243 (2025)
\bibitem{Bolton:1988af} R.~Bolton \textit{et al.}, Phys.  Rev. D \textbf{38}, 2077 (1988)
\bibitem{MEG:2020zxk} A.~Baldini \textit{et al.}, Eur. Phys. J. C  \textbf{80}, 858 (2020)
\bibitem{Knapen:2024fvh} S.~Knapen, T.~Opferkuch, D.~Redigolo and M.~Tammaro, JHEP  \textbf{6}, 189 (2025)
\bibitem{Echenard:2014lma} B.~Echenard, R.~Essig and Y.~Zhong  JHEP  \textbf{1}, 113 (2015)
\bibitem{Hesketh:2022wgw} G.~Hesketh \textit{et al.} [Mu3e], ``The Mu3e Experiment,'' arXiv:2204.00001.
\bibitem{Suzuki:1987jf} T.~Suzuki, D.~Measday, and J.~Roalsvig, Phys. Rev. C  \textbf{35}, 2212 (1987)
\bibitem{Szafron_2016} R.~Szafron and A.~Czarnecki, Physics Letters B  \textbf{753}, 61 (2016)
\bibitem{Fontes:2025mps} D.~Fontes and R.~Szafron, J. High Energy Phys. \bf{2511}, 166 (2025)
\bibitem{SINDRUMII:2006dvw} W.~H.~Bertl \textit{et al.} [SINDRUM II], Eur. Phys. J. C \textbf{47}, 337-346 (2006)
\bibitem{bartoszek2015mu2e} L.~Bartoszek \textit{et al.} [Mu2e], ``Mu2e Technical Design Report,'' arXiv:1501.05241
\bibitem{COMET:2009qeh} Y.~G.~Cui \textit{et al.} [COMET], ``Conceptual design report for experimental search for lepton flavor violating $\mu^-$ - $e^-$ conversion at sensitivity of $10^{-16}$ with a slow-extracted bunched proton beam (COMET),'' KEK-2009-10
\bibitem{Krikler:2016qij} B.~E.~Krikler, ``Sensitivity and Background Estimates for Phase-II of the COMET Experiment,'' doi:10.25560/45365
\bibitem{osti_5665919} R.~Dzhilkibaev and V.~Lobashev, Soviet Journal Of Nuclear Physics \textbf{49}, 2 (1989)
\bibitem{Hino:2014bpx} Y.~Hino \textit{et al.}, Nucl. Phys. B Proc. Suppl. \textbf{253}, 206 (2014)
\bibitem{Mu2e:2022ggl} F. Abdi \textit{et al.}, Universe \textbf{9}, 54 (2023)
\bibitem{Schechter:1981bd} J.~Schechter and J.~Valle, Phys.  Rev. D  \textbf{25}, 2951 (1982)
\bibitem{Hirsch:2006yk} M.~Hirsch, S.~Kovalenko and I.~Schmidt, Phys. Lett. B  \textbf{642}, 106 (2006)
\bibitem{SINDRUMII:1998mwd} J.~Kaulard \textit{et al.}, Phys. Lett. B \textbf{422}, 334 (1998)
\bibitem{Lee:2021hnx} M.~Lee and M.~MacKenzie, Universe \textbf{8}, 227 (2022)
\bibitem{Huang:2022xii} S.~Huang, ``Search for Lepton Flavor Violation in Two Body Muon and Pion Decay at Rest,'' FERMILAB-THESIS-2022-31
\bibitem{PIONEER:2022yag} W.~Altmannshofer \textit{et al.} [PIONEER], arXiv:2203.01981
\bibitem{Mu2e-II:2022blh} K.~Byrum \textit{et al.} [Mu2e-II], arXiv:2203.07569
\bibitem{Ball2017} M.~Ball  \textit{et al.}, ``The PIP-II Conceptual Design Report,'' doi:10.2172/1346823
\bibitem{Cirigliano:2009bz} V.~Cirigliano, R.~Kitano, Y.~Okada and P.~Tuzon, Phys. Rev. D \textbf{80}, 013002 (2009)
\bibitem{Borrel:2024ylg} L.~Borrel, D.~G.~Hitlin and S.~Middleton, Nucl. Phys. A \textbf{1062}, 123161 (2025)
\bibitem{CGroup:2022tli} M.~Aoki \textit{et al.}, arXiv:2203.08278
\bibitem{KUNO2005376} Y.~Kuno, Nucl. Phys. B Proc. Suppl. \textbf{149}, 376-378 (2005)
\bibitem{Pontecorvo:1957cp} B.~Pontecorvo, Sov. Phys. JETP \textbf{6}, 429-431 (1958)
\bibitem{Feinberg:1961zza} G.~Feinberg and S.~Weinberg, Phys. Rev. \textbf{123}, 1439-1443 (1961)
\bibitem{Heeck:2024uiz} J.~Heeck and M.~Sokhashvili, Phys. Lett. B \textbf{852}, 138621 (2024)
\bibitem{Conlin:2020veq} R.~Conlin and A.~A.~Petrov, Phys. Rev. D \textbf{102}, no.9, 095001 (2020)
\bibitem{Willmann:1998gd} L.~Willmann \textit{et al.}, Phys. Rev. Lett \textbf{82}, 49 (1999)
\bibitem{Bai:2024skk} A. Bai, \textit{et al.}, Nucl. Sci. Tech. {\bf 37}, 57 (2026)
\bibitem{Cai:2023caf} H.~Cai \textit{et al.}, Phys. Rev. Accel. Beams \textbf{27}, 023403 (2024)
\bibitem{Beare:2020gzr} J. Beare, \textit{et al.}, PTEP \textbf{2020}, no.12, 123C01 (2020)
\end{thebibliography}
\end{document}